\documentstyle[eqsecnum,aps,prb]{revtex}
\newcommand{\non}{\nonumber}

\newcommand{\ii}{{i}}
\newcommand{\be}{\begin{eqnarray}}
\newcommand{\en}{\end{eqnarray}}

\newcommand{\r}{\bbox{r}}
\newcommand{\k}{\bbox{k}}
\newcommand{\q}{\bbox{q}}
\newcommand{\bigQ}{\bbox{Q}}
\newcommand{\bv}{\bbox{v}}
\newcommand{\bb}{\bbox}
\newcommand{\ga}{{\bbox{a}^3}}

\begin{document} 
\draft

\title{Signature of the staggered flux state around a 
superconducting vortex
in underdoped cuprates}
\author{Jun-ichiro Kishine}

\address{Department of Theoretical Studies,
Institute for Molecular Science,
Okazaki 444-8585, Japan }
\address{and Department of Functional Molecular Science, 
Graduate University for Advanced Studies, Okazaki 444-8585, Japan}

\author{Patrick A. Lee, and Xiao-Gang Wen}

\address{Department of Physics,
Massachusetts Institute of Technology,
Cambridge, Massachusetts 02139}

\date{\today} 
\maketitle

\begin{abstract}

Based on 
the SU(2) lattice gauge theory formulation of the $t$-$J$ model,
we  discuss  possible
signature of the unit cell doubling associated with 
the staggered flux (SF) state in the lightly doped spin liquid.
Although the SF state appears 
only dynamically in
a uniform  $d$-wave superconducting (SC) state, 
a topological defect [SU(2) vortex] freezes  the SF state inside the vortex core.
Consequently, the unit cell doubling shows
up in the hopping ($\chi_{ij}$) and pairing ($\Delta_{ij}$) order parameters of
physical electrons.
We find that whereas the center in the vortex core is a SF state,
as one moves away from the core center,
a correlated staggered modulation of
$\chi_{ij}$ and $\Delta_{ij}$ becomes predominant.
We predict that over the region  
outside the core and inside the internal gauge field penetration depth
around a vortex center,
the local density-of-states (LDOS) exhibits staggered peak-dip (SPD) 
structure inside the V-shaped profile  when measured on the bonds.
The SPD structure has its direct origin in
the unit cell doubling associated with the SF core and the robust topological texture, which  has little to do with the symmetry of the $d$-wave
order parameter. Therefore the structure 
may survive the tunneling matrix element effects and  easily be detected by 
STM experiment. 
\end{abstract}
\vskip 0.5 in
%\begin{multicols}{2}
\section{Introduction}

\baselineskip15pt

High $T_c$ superconductors are  doped Mott insulators.
Soon after the discovery, Anderson proposed that 
the strong correlation physics of
the doped Mott insulator is well captured 
 by the $t$-$J$ model.
Taking account of competition between the hole kinetic energy $xt$
 and the spin exchange energy $J$,
 he proposed that
 the spin liquid states formed out of
  the  resonating valence bond (RVB) singlets  
 are a good starting point to study this model.\cite{Anderson87}
 A standard way of enforcing the constraint of 
 no double occupancy in
 the $t$-$J$ model is the slave boson formalism where
a physical
electron operator $c_{i\sigma}$ with spin $\sigma$ 
at the site $i$ 
is splintered into auxiliary spin-$1/2$ fermion $f_{i\sigma}$ and
charge-1 boson $b_i$: $c_{i\sigma}= f_{i\sigma} b_i^\dagger$.
One line to describe 
the spin liquid state is to start
from mean field (MF) 
decoupling,
\cite{Affleck-Marston88,Kotliar-Liu88,Suzumura-Hasegawa-Fukuyama88}
$
\Delta_{ij}=\langle \epsilon_{\sigma\bar\sigma}
f_{i\sigma}f_{i\bar\sigma}\rangle
$
and
$
\chi_{ij}=\langle f_{i\sigma}^\dagger f_{i\sigma}\rangle
$
which characterize the spin liquid state formed out of
the RVB singlets.
The phases of $\chi_{ij}$ and $\Delta_{ij}$ 
transform as the
lattice gauge fields under local U(1) transformation which naturally
leads us to  a U(1) 
gauge theory.\cite{Baskaran-Anderson88,Ioffe-Larkin89,Lee-Nagaosa92}
At zero doping,
the $t$-$J$  model reduces to an antiferromagnetic Heisenberg model
which has an exact local SU(2) gauge symmetry.\cite{Affleck-Zou-Hsu-Anderson88}
Then, the translationally invariant
solution can be described as a $\pi$-flux state\cite{Affleck-Marston88}
or a $d$-wave pairing state\cite{Kotliar-Liu88} with
$|\chi_{ij}|=|\Delta_{ij}|$. 
These apparently different mean-field ansatz  
describe exactly the same MF state, since they
are just  SU(2) gauge equivalent.

In the U(1) slave-boson formulation, however,
the SU(2)  symmetry is broken upon hole doping
due to appearance of the boson hopping term.
Consequently, the $d$-wave superconducting (SC)  state and
the flux state are no longer equivalent.
For small doping and small $J/t$,
the $\pi$-flux phase at zero doping is disfavored against
the staggered flux (SF) phase with 
$|\chi_{ij}|>|\Delta_{ij}|$.\cite{Affleck-Zou-Hsu-Anderson88,Poilblanc-Hasegawa90,Ubbens-Lee92}  
The SF state, however, breaks physical symmetries associated with
the time reversal   and
the spatial translation,
which causes the {\it unit cell doubling} and
{\it staggered orbital currents} of the physical holes.
Eventually
the SC phase is picked out as the MF solution out of a infinite number
of degenerate states upon 
doping.\cite{Kotliar-Liu88,Suzumura-Hasegawa-Fukuyama88,Fukuyama92}
However, it is still quite natural to expect the SF state is nearly degenerate 
with the SC state
in the lightly  doped  spin liquid states.
That is to say, as far as we confine ourselves to the
spin liquid state, 
the SU(2) gauge structure at zero doping may still be useful 
to describe the low energy states
in the underdoped regime, which are missing in the U(1) formulation.
To substantiate this idea, Wen and Lee\cite{Wen-Lee96,Lee-Nagaosa-Ng-Wen98}
introduced an SU(2) boson doublet $(h_{i})^T=(b_{i1}, b_{i2})$
and constructed the effective model 
which recovers  local SU(2) symmetry even upon doping.
From this viewpoint, 
the SF state plays
a crucial role to describe 
the low-energy spectrum of the lightly doped spin liquid state.
The question which we must consider next is
{\it how to detect a signature of the SF state
contained in the low-energy excitation spectrum.}
The first step in this direction
was addressed by Ivanov, Lee, and Wen\cite{Ivanov-Lee-Wen00}
who found a signature of the staggered current-current correlation
by using a Gutzwiller-projected $d$-wave pairing wavefunction.
This is naturally interpreted as a consequence of the
quantum fluctuations around the SC state
toward the SF state.
Leung\cite{Leung00} further sought for
a signature of the SF state and
found the current-current correlation in the 
$d$-wave SC state by using 
exact diagonalization of the $t$-$J$ model
for a system with two holes on a 32-site lattice.

In the experimental side,  
structure of the low-energy  excitations in the underlying
\lq\lq normal\rq\rq metallic phase
is concealed by a phase transition to bulk superconductivity.
One promising  way to escape from this situation
is to introduce 
the topological defect
into the superconducting phase, 
i.e., the vortex.
Inside the vortex core, 
low-energy properties of the normal metallic phase show up 
against the surrounding superconducting phase.
Remarkable progress in low-temperature
 STM technique with atomic resolution\cite{Pan-Hudson-Davis99} 
has given us  good opportunities
to look into the electronic states around 
the superconducting vortex.\cite{Maggio-Aprile95,Renner98,Pan00} 
Recent STM experiments 
on Bi$_2$Sr$_2$CaCu$_2$O$_8$ (BSCCO)\cite{Renner98,Pan00}
revealed the striking fact that the
normal core electronic state exhibits the
\lq\lq pseudogap\rq\rq structure
characteristic of the normal state pseudogap
above $T_c$.
A description of a vortex core based on 
conventional BCS theory requires that the superconducting order parameter vanishes inside the core,
which is usually accompanied by the vanishing of the energy gap.
The experimental finding thus strongly suggests 
that the electronic structure of the vortex core
is qualitatively different from that given by conventional picture.

The theoretical description of the normal core
in the light of the strong correlation physics, however,  remains
unresolved.\cite{Nagaosa-Lee92,Sachdev92,Himeda-Ogata-Tanaka-Kashiwaya97,Tsuchiura-Tanaka-Ogata-Kashiwaya99,Franz-Tesanovic00,Han-Wang-Lee00,Wang-Han-Lee01}
In the SU(2) picture, since the SF state is nearly degenerate with the SC state,
it is naturally expected that by frustrating the SC state the SF state
will be revealed inside the core.
Based on this idea, Lee and
Wen\cite{Lee-Wen00} proposed a model of the vortex with a SF
core, characterized by a pseudo gap and  staggered orbital current.
Quite recently,
Han, Wang, and Lee  found evidence of the SF order near 
the vortex core
by using Gutzwiller projected U(1) slave-boson mean-field 
wave function.\cite{Han-Wang-Lee00,Wang-Han-Lee01}
These numerical 
results so far\cite{Ivanov-Lee-Wen00,Leung00,Han-Wang-Lee00,Wang-Han-Lee01} 
strongly suggest that the
SF state is a key ingredient in the $t$-$J$ model.
 
The  vortex with the SF core [SU(2) vortex]  offers us an opportunity to
experimentally detect 
the SF state at low temperatures below $T_c$, whereas  
it may be difficult to probe the staggered
current pattern 
in the zero-field uniform SC state 
because of spatial and temporal fluctuations.
Possible experimental tests of the SF core 
were  proposed as summarized below.\cite{Lee-Wen00}
(1) Cyclotron
resonance or Shubnikov-de Haas experiments   
in a high quality underdoped sample at $H>H_{c2}$
can detect  the  small Fermi pockets around $(\pm\pi/2,\pm\pi/2)$ points
with not uniformly spaced 
Landau levels. 
(2) $\mu$-SR or neutron scattering experiments 
can directly detect
the staggered currents which   produce a small staggered
magnetic field of order
 $10$ gauss.\cite{Hsu-Marston-Affleck91}
Intensity of the signal may 
increase upon increasing  $H$, since the increasing $H$
excites more vortices with the core size being independent of $H$.
(3) NMR experiments can detect
side bands in the Y NMR line in YBCO samples
with a splitting independent of $H$ 
but with weight proportional to
$H$. For this purpose,
Y$_2$Ba$_4$Cu$_7$O$_{15}$ may be ideal, because  there are
asymmetric bi-layers where the Y ion sits in between,
and it may be possible to have
one plane of the bi-layer optimally
doped while the other plane (next to the double chain) remains underdoped, i.e.,
the staggered magnetic field at the Y site  does not cancel.

Now we are naturally lead to  the following question:
is it possible to detect a signature of the unit cell doubling
associated with the SF core
through the state-of-the-art STM technique?
It turned out that there is no effect inside the SF core, because
what is staggering in the SF state 
is  the currents, which
 does not show up in the charge density.
This situation motivated us to look at the region outside the core.
We addressed this problem in our previous paper\cite{Kishine-Lee-Wen01} 
and found  that whereas the center in the vortex core is a SF state, 
as one moves away from the core center, 
a correlated staggered modulation of the hopping amplitude $\chi_{ij}$
and pairing amplitude $\Delta_{ij}$
of the {\it physical} electrons  becomes predominant.
We predicted that in this region,
the LDOS exhibits staggered modulation when measured 
on the bonds, which may be
directly detected by STM experiments.

In this paper, we give a full account of the results summarized in 
Ref.[\cite{Kishine-Lee-Wen01}]
and examine the LDOS around the SU(2) vortex in detail.
The outline  is as follows.
In Sec.~II, we will give an overview of the SU(2) lattice gauge 
theory formulation
of the $t$-$J$ model (Sec.~II~A) and then
discuss the topological texture of the SU(2) boson condensate based on 
the O(4) $\sigma$-model (Sec.~II~B).
We are mainly 
concerned with the LDOS outside the core through which 
we detect the unit cell doubling   stabilized by
the robust topological texture.
For this purpose, a close study of the vortex core state
is not necessary.
To take account of the phase winding, we will apply
a simple London model for a single 
vortex to the SU(2) vortex model (Sec.~II~C).
 In Sec.~III, we discuss the hopping and pairing order parameters of
 the {\it physical} electron around the vortex. For this purpose,
 we perform
  an appropriate local SU(2) gauge transformation (Sec.~III~A).
  Then, we argue in detail that 
as one moves away from the core center, 
a correlated staggered modulation of $\chi_{ij}$ and $\Delta_{ij}$  
becomes predominant (Sec. III B). 
In Sec. IV, we evaluate 
the LDOS outside the core.
Formulation of the LDOS at an arbitrary point on lattice is given in Sec. IV A.
It is demonstrated that the LDOS  exhibits conspicuous staggered pattern 
only when measured on the bonds.
To obtain the LDOS, we compute the
 lattice propagator by using
two complementary approaches, which are presented in Sec. IV B and Sec. IV C.
Finally, concluding remarks are given 
 in Sec. V.

\section{SU(2) vortex with the staggered flux core}
In this section,
we recapitulate the SU(2) lattice gauge  theory formulation
of the $t$-$J$ model and then
discuss  the SU(2) vortex model in some detail.

\subsection{SU(2) lattice gauge theory formulation of the $t$-$J$ model}

The $t$-$J$ model Hamiltonian is given by
\be
H=-t\sum_{<i,j>,\sigma} \left( c_{i\sigma}^\dagger c_{j\sigma}+{\rm H.c.}\right)
+J\sum_{<i,j>}\left({\bbox{S}}_i\cdot {\bbox{S}}_j-{1\over 4}n_i n_j,
\right)\label{t-J}
\en
where $c_{i\sigma}^\dagger$ and  $c_{i\sigma}$
are the projected electron operators with the constraint
$n_i\leq 1$.
In the SU(2) slave-boson approach,\cite{Wen-Lee96,Lee-Nagaosa-Ng-Wen98}
a physical electron  is represented as
an SU(2) singlet formed out of
the \lq\lq isospin\rq\rq doublets of 
the fermion ($\psi_{i\sigma}$) and boson ($h_i$):
\be
c_{i\sigma}={1\over\sqrt{2}}h_{i}^{\dagger}\psi_{i\sigma}
={1\over\sqrt{2}}
(b_{i1}^{\dagger}f_{i\sigma}+\epsilon_{\sigma\bar\sigma}
b_{i2}^{\dagger}f_{i\bar\sigma}^\dagger),\label{physelec}
\en
with
\be
\psi_{i\sigma}=\pmatrix{
f_{i\sigma}\cr
\epsilon_{\sigma\bar\sigma} f_{i\bar\sigma}^\dagger},\,\,\,\,
h_{i}=\pmatrix{b_{i1}\cr b_{i2}}.
\en
The physical hole density
$\langle b_{i1}^\dagger b_{i1}+ b_{i2}^\dagger b_{i2}\rangle=x$
is enforced by the chemical potential $\mu$.
We need to introduce
the temporal component of the gauge field ${\bbox a}_{0i}$ to
ensure the
projection of the Hilbert space
onto the SU(2) singlet subspace,
$
({1\over 2}\psi_{i\sigma}^\dagger \bb{\tau}\psi_{i\sigma}
+h_{i}^\dagger\bb{\tau} h_{i})|{\rm phys}\!\!>=0
$,
which is identical to that of the original $t$-$J$ model.
The conventional
 U(1) slave-boson
 $b_i$
 is now regarded as  the SU(2) boson doublet having only its
 isospin \lq\lq up\rq\rq  component:
$(h_{i}^{(0)})^T=(b_{i}, 0)$.
The spin liquid state is characterized by
the order parameters
$
\Delta_{ij}=\langle \epsilon_{\sigma\bar\sigma}
f_{i\sigma}f_{j\bar\sigma}\rangle
$
and
$
\chi_{ij}=\langle f_{i\sigma}^\dagger f_{j\sigma}\rangle
$ which
constitute a $2\times2$ matrix
\be
U_{ij}=
\pmatrix{-\chi_{ij}^\ast&\Delta_{ij}\cr 
\Delta_{ij}^\ast&\chi_{ij}}.
\en
By this decoupling, the spin-exchange term is replaced with
$
{\bbox{S}}_i\cdot {\bbox{S}}_j\to
{3J\over 16}\sum_{\sigma}\psi_{i\sigma}^\dagger U_{ij}\psi_{j\sigma}
+{3 J\over 16}{\rm Tr}\left[U_{ij}^\dagger
U_{ij}\right].
$
We should stress here that in the presence of the $b_2$-boson
$\chi_{ij}$ and $\Delta_{ij}$ cannot be
interpreted as the hopping and pairing order parameters of 
a {\it physical electron} [see Eq.~(\ref{physelec})].
The \lq\lq phase\rq\rq  of $U_{ij}$ is now interpreted as the
SU(2) lattice gauge fields:\cite{Affleck-Zou-Hsu-Anderson88,Dagotto-Fradkin-Moreo88}
\be
\bar U_{ij}=U_{ij} 
\exp[-i\bb{a}_{ij}\cdot\bb{\tau}],
\en
where $\bb{\tau}=(\tau^1,\tau^2,\tau^3)$ are Pauli matrices
and $\bb{a}_{ij}=({a}_{ij}^1,{a}_{ij}^2,{a}_{ij}^3)$ is the gauge field on
every link.
Now the $t$-$J$ model is described by
the fermion-boson system interacting
with the SU(2) lattice gauge 
field,\cite{Wen-Lee96,Lee-Nagaosa-Ng-Wen98}
described by the Lagrangian
:
$
L_0=L_0^{\rm F}+L_0^{B}+{\tilde J\over 2}\sum_{<i,j>}
{\rm Tr}[\bar U_{ij}^\dagger \bar U_{ij}]
$
with
\be
L_0^{\rm F}&=&
{1\over 2}\sum_{i,j,\sigma}
\psi_{i\sigma}^\dagger 
\left[\delta_{ij}\partial_\tau +\tilde J\bar U_{ij}\right] 
\psi_{j\sigma}
%\non\\&&\,\,\,\,\,\,\,\,\,\,\,\,\,\,\,\,
+
{1\over 2} \sum_{i,\sigma}\psi_{i\sigma}^\dagger i\bb{a}_{0i}\cdot
\bb{\tau}
 \psi_{i\sigma},\label{fermion}\\
L_0^{\rm B}&=&\sum_{i,j}
h_{i}^\dagger \left[\delta_{ij}(\partial_\tau-\mu) +
\tilde t \bar U_{ij}\right]  h_{j}
%\non\\&&\,\,\,\,\,\,\,\,\,\,\,\,\,\,\,\,
+\sum_{i}
h_{i}^\dagger i{\bb{a}}_{0i}\cdot \bb{\tau} h_{i}
,\label{boson}
\en
where $\tilde J=3J/8$ and $\tilde t=t/2$.
The mean-field solution 
is obtained by 
integrating out the fermions and
minimizing the mean-field energy $E(\{ U_{ij}, h_i\})$,
which leads to 
$U_{ij}$ on the links and 
the boson $h_i$ on the sites.

The SU(2) gauge invariance is realized through the relation
$ E(\{ \bar U_{ij}, h_i\}) = E(\{ W_i\bar U_{ij}W_j^\dagger, W_i h_i\})$,
for any
$W_i \in {\rm SU(2)}$.
Thanks to the SU(2) symmetry,
we can choose a convenient gauge fixing 
to describe the MF state in an SU(2) invariant way.
Convenient gauge choices in the underdoped regime are
the \lq\lq $d$-wave  gauge\rq\rq
or the \lq\lq staggered-flux (SF) gauge\rq\rq
specified by
\be
U_{ij}^d&=&-\chi_0 \tau^3+(-1)^{i_y+j_y}\Delta_0\tau^1,\label{dan}\\
U_{ij}^{\rm SF}
&=&-A
\tau^3
\exp[\ii(-1)^{i_x+j_y}\Phi_0\tau^3],\label{sfan}
\en
respectively,
where 
%$\chi_0=|\chi_{ij}|$, $\Delta_0=|\Delta_{ij}|$,
$A=\sqrt{\chi_0^2+\Delta_0^2}$ and
$\Phi_0=\tan^{-1}(\Delta_0/\chi_0)$.
Eq.~(\ref{dan}) describes fermions 
with $d$-wave pairing order parameters,
while
Eq.~(\ref{sfan}) describes fermions hopping with flux $\pm 4 \Phi_0$
on alternating plaquettes.\cite{Affleck-Marston88}
At zero doping ($x=0$) there is no boson and
these apparently different mean-field ansaz  
describe exactly the same MF state, since 
$U_{ij}^d$ and $U_{ij}^{\rm SF}$
are  just  SU(2) gauge equivalent, i.e.,
$U_{ij}^d=w_i U_{ij}^{\rm SF}w_j^\dagger$
[$E(\{U_{ij}^d\})=E(\{w_i U_{ij}^{\rm SF}w_j^\dagger\})
=E(\{U_{ij}^{\rm SF}\})$], where
the transformation is explicitly given by
\be
w_i=\exp[{\ii(-1)^{i_x+i_y}{\pi\over 4}\tau^1}].\label{dSFconverter}
\en
Upon doping, however,
the $U_{ij}^d$ with the U(1) 
boson condensate $(h_{0i})^T=(b_i,0)=(\sqrt{x},0)$
characterizes the physical $d$-wave SC state, while 
the $U_{ij}^{\rm SF}$ with the U(1) boson 
condensate $(h_{0i})^T=(\sqrt{x},0)$
characterizes the physical SF state.
These states are no longer physically equivalent because of
the presence of the boson condensate 
[$E(\{U_{ij}^d,h_{0i}\})\neq E(\{U_{ij}^{\rm SF},h_{0i}\})$]
 and 
the SC phase is picked out as the MF 
solution.\cite{Poilblanc-Hasegawa90,Ubbens-Lee92}
Accordingly,
the \lq\lq flux\rq\rq  $\Phi_0=\tan^{-1}(\Delta_0/\chi_0)$
decreases from
$\Phi_0=\pi/4$ ($\pi$-flux phase) 
upon doping.\cite{Poilblanc-Hasegawa90,Ubbens-Lee92}

The advantage of the SF gauge is that
it is apparent that the SU(2) symmetry
has been broken down to the residual
U(1), which we denote U(1)$_{\rm res}$,
 since $U_{ij}^{\rm SF}$
contains only $\tau^3$.\cite{Wen91}
The lattice gauge fields $a_{ij}^1$ and $a_{ij}^2$
become massive by the Anderson-Higgs mechanism and can be ignored, while
 $a_{ij}^3$ remains massless and is the important low
energy degrees of
freedom which should be included, i.e. we consider
\be
\bar U_{ij}^{\rm SF}
=-A\tau^3
\exp\left[
i (-1)^{i_x+j_y}\Phi_0\tau^3\right]
\exp\left[-i a_{ij}^3\tau^3\right].\label{Ua}
\en
In this gauge, we can discuss a vortex structure under the external
magnetic field
in a way quite similar to
the conventional BCS vortex where the gauge structure is characterized by 
only
the electromagnetic (EM) U(1)$_{\rm em}$. The difference is that,
in our problem,
the gauge structure is characterized by U(1)$_{\rm em}\otimes$U(1)$_{\rm 
res}$.

\subsection{O(4) $\sigma$-model description of the local boson condensate}
In the presence of a magnetic field, the mean-field solution contains 
vortices.
The SU(2) vortex model\cite{Lee-Wen00} was discussed based on
the O(4) $\sigma$-model description for a slowly varying boson
condensate.\cite{Lee-Nagaosa-Ng-Wen98}
The basic idea is that at low temperatures the bosons are nearly 
condensed
to the bottom of the band
and are slowly varying in space and time.
The ansatzs (\ref{dan}) and (\ref{sfan})
gives the one-boson dispersion
$\xi^{\rm B}_{\k}=-\tilde t A
(\cos ^2 k_x+\cos^2 k_y+2\cos 2\Phi_0\cos
k_x\cos k_y)^{1/2}$.
The $b_1$ and $b_2$ bosons  
 are then nearly condensed
to the band bottom  $(0,0)$ and $(\pi,\pi)$, 
respectively.\cite{comment-on-gauge}
On the other hand, the fermions are fluctuating over the lattice scale and
can be integrated out,
after choosing an $\bb{a}_{0i}$ field which minimize the action locally.
This view is  in the spirit of the Born-Oppenheimer
approximation.\cite{Lee-Nagaosa-Ng-Wen98}
In the SF gauge given by Eq.~(\ref{sfan}),
the local boson condensate (LBC)
can be written as
\be
\bar h_{i}^{\rm SF}
=\sqrt{x}\pmatrix{z_{i1}\cr-i(-1)^{i_x+i_y} z_{i2}},
\label{CP1}
\en
where $z_{i1}$ and $z_{i2}$ [CP$^1$ fields]
 are slowly varying  in space and time and
parameterized by
\be
z_{i1}=e^{\ii\varphi_{1i}}\cos{\theta_i\over 2},\,\,\,\,\,
z_{i2}=e^{\ii\varphi_{2i}}\sin{\theta_i\over 2},
\label{CP1rep}
\en
with the internal phases being given by
\be
\varphi_{1i}=\alpha_i-\phi_i/2,\,\,\,\,\,
\varphi_{2i}=\alpha_i+\phi_i/2.
\en
We shall give some remarks on the expression Eq.~(\ref{CP1}) in appendix A.

The overall  phase angle  $\alpha$ is associated with
the  U(1)$_{\rm em}$.
The internal SU(2) gauge symmetry is broken down to U(1)$_{\rm res}$
and
the angles $\phi$ and $\theta$ are interpreted as
polar angles of the manifold of the LBC: 
SU(2)/U(1)$_{\rm res}\simeq$ S$^2$.
Topological stability of vortex formation is
indicated by the non-trivial topology,
$\pi_2[$SU(2)/U(1)$_{\rm res}]=\pi_1[$U(1)$_{\rm res}]=\bf{Z}$.
The internal degrees of freedom of the LBC
is visualized by the vector
\be
{\bbox I}_i
=z^\dagger_i{\bbox\tau} z_i
=
(\sin\theta_i\cos\phi_i,\sin\theta_i\sin\phi_i,\cos\theta_i),
\en
which has the meaning of the quantization axis
for the $z$ fields,
$
(z_i)^T=(z_{i1}, z_{i2}).
$
In the SF gauge, the uniform $d$-wave SC state and the uniform SF state 
are described by $\theta_i=\pi/2$ and $\theta_i=0,\pi$, 
respectively.
The  angle $\phi_i$ is associated with  the residual gauge symmetry
U(1)$_{\rm res}$ which
is further broken down to $\{\bb{0}\}$ upon the bose condensation
which triggers the superconducting phase transition.

%%%%%%%%%%%%%%%%%%%%%%%%%%%%%%%%%%%%%%%%
%%%%%%%%%%%%%%%%%%%%%%%%%%%%%%%%%%%%%%%%
The low energy dynamics of the LBC is described by
an anisotropic O(4) $\sigma$-model coupled to the gauge 
fields.\cite{Lee-Nagaosa-Ng-Wen98}
Since we are only concerned with static configurations,
we shall ignore the time dependent terms from now on. 
The free energy associated with this model is
 written in a form
$F_{\rm eff}=F_{\rm K}+F_\perp+F_{A}+F_{a}$, explained below.
In the SU(2) formulation, only the boson can carry charge.
Under the magnetic field, the  boson hopping-pairing matrix in
Eq.~(\ref{boson})
acquires an EM
Peierls phase: $\bar U_{ij}^{\rm SF}\to \bar U_{ij}^{\rm SF} \exp[i{e\over c}
\int_{\r_i}^{\r_j}\bb{A(\r)}\cdot d\r]$.
%By noting
%$\tilde t\sum_{<i,j>}\bar h_{i}^{{\rm SF}\dagger}\bar U_{ij}^{\rm SF}
%e^{i{e\over c}\int_{\r_i}^{\r_j}\bb{A(\r')}\cdot d\r'} \bar h_{j}^{\rm SF}
%=x\tilde t \sum_{<i,j>}z_{i}^{\dagger}\tau^3 \bar U_{ij}^{\rm SF}e^{i{e\over c}
%\int_{\r_i}^{\r_j}\bb{A(\r')}\cdot d\r'}z_{j}
%$
%and
Taking a continuum limit,
the kinetic part is written as
\be
F_{\rm K}={x\over 2m_b}\int d\r
|\bb{\cal D}
z|^2,\label{FK}
\en
where we introduced the boson mass $m_b\sim 1/t$.
The covariant derivative is given by
$
\bb{\cal D}=\bb{\nabla}+i\bb{a}^3\tau^3-i{e\over c}\bb{A},
$
where
we introduced the continuum limit of the $a_{ij}^3$ field through
$
a_{ij}^3
=(\r_i-\r_j)\cdot\bb{a}^3({\r_i+\r_j\over 2}).
$

The anisotropy term is phenomenologically given in a form
\be
F_\perp={x^2\tilde J\over 2}
\int d\r
\left[{4\over c_1}| z_1 z_2|^2
+{1\over c_3}(|z_1|^2-|z_2|^2)^2\right],\label{Fperp}
\en
with $c_1$ and $c_3$ being
numerical constants of order of unity.\cite{Lee-Nagaosa-Ng-Wen98}
This term describes energy cost associated with small fluctuations
of the LBC around the SC state ($\theta=\pi/2$).
For $c_3< c_1$, the $\bb{I}$-vector prefers to lie in the 
$\tau_1$-$\tau_2$ plane (equatorial plane) and the SC state is favored.

The conventional EM Maxwell term  is given by
\be
F_A={1\over 8\pi}\int d\r(\bb{\nabla}\times \bb{A})^2.\label{FA}
\en
The fourth term $F_a$, 
the internal gauge field kinetic term,
is dynamically induced by
integrating out the fermion degrees of freedom 
although we have no such a term initially
at the relevant highest energy scales of the fermions $\sim \chi_0 \tilde J$. 
We have
\be
F_a=
{\sigma\over 2}\sum_{\bb{q}}\sum_{\mu,\nu=x,y}
a^3_\mu(\bb{q}) \Pi_{\mu\nu}^{\rm F}(\bb{q})
a^3_\nu(\bb{q}),\label{Fgauge}
\en
where $\sigma=\sqrt{\tilde J\Delta}$
and the fermion polarization bubble originating from
the  coupling term of the Dirac fermion current and gauge field 
%$\bar\psi\gamma^\mu a_\mu^3\tau^3\psi$ ($\gamma^\mu$ is
%properly constructed gamma matricies associated with the Dirac
%fermions)
is given by
\be
\Pi_{\mu\nu}^{\rm F}(\bb{q}) =\left(\delta_{\mu\nu}-{q_\mu q_\nu\over q^2}
\right)|\bb{q}|.\label{gfct}
\en
We note that this does not take the EM Maxwell form which is proportional 
to
$\bb{q}^2$, and consequently gives rise to a non-local kernel in real 
space:
\be
F_a={\sigma\over 2}\int d\r\int
d\r'\kappa(\r-\r')\bb{h}(\r)\cdot\bb{h}(\r'),\label{Fareal}
\en
where $\bb{h}(\r)=\bb{\nabla}\times \bb{a}^3(\r)$ and
$
\kappa(\r-\r')
=\sum_{\bb{q}} e^{-i{\bb{q}}\cdot(\r-\r') } \kappa_{\bbox{q}}
$ 
with
\be
\kappa_{\q}=1/|\bb{q}|,
\label{ftk}
\en
instead of $\kappa_{\q}=1$
in case of the conventional EM kernel.

\subsection{London model of a single SU(2) vortex}
In the model of the vortex proposed by Lee and Wen,
both $\alpha$ and $\phi/2$ wind by
$\pi$ and consequently give
an appropriate $hc/2e$ vortex for
the EM
gauge field ${\bbox A}({\bbox r})$.
This way  of winding is specified by
\be
{\bbox \nabla} \alpha={\bbox \nabla}
{ \phi\over 2}={\hat{\bf e}_\phi\over2r}
\en
which lead to
$
{\bbox \nabla} \varphi_1=0
$
and
$
{\bbox \nabla} \varphi_2={\hat{\bf e}_\phi
}/r
$,
where $\hat{\bf e}_\phi$ denotes the azimuthal unit vector in the 
physical
space.
That is to say, only
$b_2$ changes its phase
$\varphi_2$
by $2\pi$ as we go around the vortex,
while
$b_1$ does not.

The texture of the $\bb{I}$-vector 
in the SF gauge
is indicated in Fig.~1(a).
In the SC state  outside the core,
${\bbox I}_i
=(\cos\phi_i,\sin\phi_i,0)$,
while
as we approach the core
$|b_2|$ must vanish and the vortex center
is represented by ${\bbox I}_i=(0,0,1)$ which is just the SF state.
The ${\bbox I}_i$-vector tilts smoothly from the equator to
the north pole as the core is approached
with a length scale denoted by $\ell_c$ which is identified with
the core size.
To determine the SU(2) vortex structure, we shall use
the \lq\lq London model\rq\rq prescription
of a single vortex
in  extremely type II BCS superconductor.\cite{Fetter-Hohenberg69}
Detailed account of the analysis is given in appendix B.
%%%%%%%%%%%%%%%%%%%%
%\noindent Fig.1
\begin{figure}
%\epsfxsize=3.0in
%\centerline{\epsffile{Fig01.eps}}
\smallskip
\caption{
(a) The texture of the $\bb{I}$-vector in the SU(2) vortex configuration
in the SF gauge.
At the center of the vortex,
${\bb I}_i$ points toward the north pole corresponding to the  SF state.
The shaded circle depicts the vortex core.
The local gauge  transformation $g_i$ transforms this configuration
to (b) in the $d$-wave gauge, where the internal phases of the bose
condensate are gauged away.
}
\end{figure}
%                 %
%%%%%%%%%%%%%%%%%%%

Although
 quantitative estimation of
 $\ell_c$ and  $\lambda_a$ is beyond the 
present simple London model analysis,
$\ell_c$ presumably extends
 over a  fermion coherence length
$\xi_F\sim v_F/\Delta$ which may amount to
a few lattice scales as suggested numerically.\cite{Han-Wang-Lee00} 
%It should also be remarked that
%we cannot explicitly analyze the electronic states inside the core.
%For example,  it is beyond the present scheme to make clear
%the existence or absence of
%the core bound state or the shape of the core boundary.
% To inquire further into the matter would lead us to
% rather specialized problem 
%which itself has provoked a great deal of controversy
%from both experimental\cite{Maggio-Aprile95,Renner98,Pan00} 
%and 
%theoretical\cite{Himeda-Ogata-Tanaka-Kashiwaya97,Soininen-Kallin-Berl%insky94,Wang-MacDonald95,Franz-Ichioka97,Morita-Kohmoto-Maki97,Yasui-%Kita99} sides. 
We here just remark 
that
 there are two kinds of vortices, because the
$\bbox{I}$-vector can also point toward  the south pole at the
vortex core:
$\theta_i=\pi$ in Eq.~(\ref{CP1rep}).  
This just expresses the state with the
staggered flux shifted by one unit cell.
If the center of the vortex is in the center of the plaquette, the degeneracy between these two kinds of vortices is broken by the circulation of the 
EM superfluid current. This is the situation considered by
Wang, Han, and Lee\cite{Wang-Han-Lee01}
in their numerical local U(1) mean field approach.
On the other hand, if the center of the vortex is on a lattice site,
the degeneracy remains and there is quantum mechanical tunneling between the two states. The tunneling rate depends on $\ell_c$ and is difficult to estimate. However, the dissipation due to quasi-particles
may suppress the tunneling rate due 
to the {\it orthogonality catastrophe}.  
Whether the two states are degenerate or not depends on short distance physics which is outside the domain of our long wavelength theory.

\section{Hopping and pairing order parameters of
the physical electrons around 
a single vortex}
\subsection{Gauge transformation of the local boson condensate}

Now that the SU(2) vortex model has been established,
we shall discuss the effects of  
the unit cell doubling and the phase winding
on the hopping and pairing order parameters of the {\it physical} electrons
around a single vortex. 
For this purpose, it is
 best to work with  the $d$-wave gauge
after
making a local gauge transformation by
\be
g_i=\exp\left[{\ii(-1)^{i_x+i_y}{\theta_i\over 2}\tau^1}\right]
\exp\left[{\ii{\phi_i\over 2}\tau^3}\right].\label{eg}
\en
The LBC is then transformed  to
\be
\bar h_{i}^{\rm SF}
\to
\bar h_{i}^{d}=g_i \bar h_{i}^{\rm SF}
=e^{i\alpha_i}
\pmatrix{\sqrt{x}\cr 0} ,
\en i.e. the $\bb{I}$-vector  points toward the north pole
{\it everywhere on lattice}, as shown in
Fig.~1(b). We here  consider only the case of a single vortex.
The  great advantage 
of the $d$-wave gauge is that the physical electron operator
is simply written as
\be
c_{i\sigma}={1\over\sqrt{2}}\bar h_{i}^{d\dagger}\psi_{i\sigma}
=e^{-i\alpha_i}\sqrt{x\over 2}f_{i\sigma}\label{physhole},
\en
i.e., the fermions behave as physical electron.
After the local gauge transformation to the $d$-wave gauge, we find
\be
&&\bar U_{ij}^{\rm SF}\to\bar U_{ij}^d=g_i \bar U_{ij}^{\rm SF}
g_j^\dagger\non\\
&=&-\tilde\chi_{ij}
\left(\tau^3 \cos{\theta_i-\theta_j\over 2}
+\tau^2 \sin{\theta_i-\theta_j\over 2}
\right)\label{Uij}
\\
&-&\tilde\Delta_{ij}\left[
\ii(-1)^{i_x+j_y}
\cos{\theta_i+\theta_j\over 2}
-(-1)^{i_y+j_y}\tau^1
\sin{\theta_i+\theta_j\over 2}\right], \non
\en
where
\be
\tilde\chi_{ij}&=& A\cos\Phi_{ij},\,\,\,\,\,\,\,\,
\tilde\Delta_{ij}=A\sin\Phi_{ij},\label{modchiDelta}\\
&& \Phi_{ij}=\Phi_0+
{(-1)^{i_x+j_y}}
v_{ij},\label{Phiij}
\en
and
\be
v_{ij}
=
{\phi_i-\phi_j\over2}
-a_{ij}^3.
\en
As an important consequence of the local gauge transformation, 
the gauge invariant quantity
$
v_{ij}
$
enters Eq.~(\ref{Uij}). 
This quantity has a meaning of
the fermion \lq\lq superfluid velocity\rq\rq 
associated with the internal gauge field $\bb{a}^3$
which is circulating around the vortex 
center [see Eq.~(\ref{va})].
For example, let us  consider 
$v_y(\r)$ along the line, 
$i_y=1/2$, assuming that the vortex center sits at $(1/2,1/2)$.
Fourier transform of Eq.~(\ref{vq}),
gives
\be
v_y(\r)={1\over 2r}-{1\over 2}\int_0^\infty dq {J_1(q i_x)\over 1+\lambda_a q},
\label{configv}
\en
where $\r=(i_x,1/2)$.
In Fig.~2, we show  the spatial distribution of 
$v_y(\r)$ by assuming the gauge field penetration depth
to be $\lambda_a=10$ with lattice unit.
We see that $v_y(\r)$  decays over the length scale $\lambda_a$,
as is naturally expected for the superfluid velocity.
%%%%%%%%%%%%%%%%%%%%
%\noindent Fig.2
\begin{figure}
%\epsfxsize=2.8in
%\centerline{\epsffile{Fig02.eps}}
\smallskip
\caption{
Spatial distribution of the superfluid velocity
$v_y(\r)$ associated with the internal gauge field $\bb{a}^3$.
We assumed the gauge field penetration depth
to be $\lambda_a=10$ with lattice unit.
%In the inset are shown the direction and magnitude of
%$\bb{v}(\r)$ along the $x$-axis. 
The origin should not be
taken too literally, since $\bb{v}$ is defined in the continuum limit.}
\end{figure}
%                 %
%%%%%%%%%%%%%%%%%%%

Let us write 
$\bar U_{ij}^d$
in the form
\be
\bar U_{ij}^d=\pmatrix{-\bar\chi_{ij}^\ast&\bar\Delta_{ij}\cr
\bar\Delta_{ij}^\ast&\bar\chi_{ij}}.\label{Udslow}
\en
%where
%\be
%\bar\chi_{ij}&=&\chi_{ij}-i(-1)^{i_x+j_y}\Delta_{ij}\cos\theta_i,\lab%el{chislow}\\
%&&\bar\Delta_{ij}=(-1)^{i_y+j_y}\Delta_{ij}\sin\theta_i\label{Delslow%}.
%\en
%It should be noted, however, 
%this expression becomes obviously invalid inside the vortex core wher%e
%the variation of the $\theta$-angle over the lattice scales may
%become important, especially in the radial direction with respect to %the vortex center. 
%This point is taken up shortly later.
An essential point is that in the $d$-wave gauge
{\it $\bar\chi_{ij}$ and $\bar\Delta_{ij}$ have the meaning of
the hopping and pairing order parameters of the physical electron},
since the physical electron operator $c_{i\sigma}$ is just proportional to the
auxiliary fermion operator $f_{i\sigma}$
[Eq.~(\ref{physhole})].
Below we discuss the meaning of $\bar\chi_{ij}$ and $\bar\Delta_{ij}$
at different limits.

\subsection{Hopping-pairing order parameters in the
vicinity and outside of the vortex core}
\subsubsection{Vicinity of the  vortex center}
First, we consider the vicinity of the  vortex center,
where
$\theta_i\sim\theta_j\sim 0$
 and Eq.~(\ref{Udslow}) becomes
\be
\bar U_{ij}^d
\sim
-A\tau^3
\exp\left[
i (-1)^{i_x+j_y}\Phi_{ij}
\tau^3\right],
\en
i.e.,
\be
\bar\chi_{ij}&=&
%\chi_{ij}+i(-1)^{i_x+j_y}\Delta_{ij}
Ae^{i(-1)^{i_x+j_y}\Phi_{ij}},\label{cijsf}\\
\bar\Delta_{ij}&=&0.\label{dijsf}
\en
%Furthermore, Eq.~(\ref{optad})
%gives
%\be
%{\bb{a}}_{0i}^d=(0,0,{x\tilde J\over c_3})
%\en
%which simply points toward the north pole.
Eq.~(\ref{dijsf}) indicates that 
the superconducting order parameter is  killed at the vortex center.
In this region,  as is directly seen from
Eq.~(\ref{cijsf}),
what is modulated is
the phase of the fermion hopping parameter which is just 
regarded as the electron hopping parameter.
We see that the sum of the phase around an elementary plaquette yields
 modulated net flux
 $\pm4\Phi_{0}+\phi_{\rm gauge}(r)$
with $\pm$ signs alternating from plaquette to plaquette.
We here introduced a gauge flux penetrating an elementary plaquette
centered at ${\bbox r}\neq 0$:
\be
\phi_{\rm gauge}(r)=\oint_{\Box}\bb{\nabla}\times\bb{v}\cdot d\bb{\ell}
\sim -{c_0^2 \over \hbar} h(r),
\en
where we retained
 the lattice constant $c_0$.
 The internal gauge field strength 
 $\bb{h}(r)=\bb{\nabla}\times\bb{a}^3$
 is given by Eq.~(\ref{gauge-field-strength}).
We see  $\phi_{\rm gauge}(r)\ll 4\Phi_{0}\sim {\cal O}(1)$ [for example,
 $\phi(2 c_0)
 =0.03$ if we take $\lambda_a=10 c_0$].
This situation just indicate the fact that the net flux is dominated by
the original staggered flux $4\Phi_0$.
Thus,
inside the core
 the staggered  phase modulation becomes predominant:
 $
\bar U_{ij}^d
\sim
-A\tau^3
\exp\left[
i (-1)^{i_x+j_y}\Phi_{0}
\tau^3\right]$.

The $\bar U_{ij}^d$ then breaks not only
 the translational symmetry
($\bar U_{ij}^d\neq \bar U_{i+\hat{\bf e}_\mu,j+\hat{\bf e}_\nu}^d$,
where $\hat{\bf{e}}_\mu$ with
$\mu,\nu=x, y$
denotes a unit vector connecting the neighboring 
sites
),
but also
 the time reversal symmetry with  respect to the local bonds
($\bar U_{ij}^d \neq [\bar U_{ij}^{d}]^\ast$).
Although 
we cannot explicitly analyze the electronic states inside the core,
 the time reversal symmetry breaking implies that the staggered fermion 
 currents
 flow 
on the bonds just as in the case of a uniform SF state. 
Once the bosons are condensed, the currents come up as
the staggered orbital currents of the  physical hole.\cite{Lee-Wen00}

In this paper, we are concerned with the possibility of
detecting a signature of the unit cell doubling
through  STM measurement.
We immediately see that there is no hope 
in the SF state, because
what is staggering in the SF phase 
is  the on-bond  currents caused by the  staggered phase 
[Eq.~(\ref{cijsf})].
Consequently,
the period doubling of the current never shows up in
 the LDOS.\cite{comment1}
This situation motivates us to look at the region outside the core.

\subsubsection{Outside the SF core}
We consider the region outside the SF core.
We approximately set
$\theta_i\sim\theta_j\sim\pi/2$ which
gives
\be
\bar U_{ij}^d&\sim&-\tilde\chi_{ij}
\tau^3+(-1)^{i_y+j_y}\tilde\Delta_{ij}
\tau^1
,
\en
i.e.
\be
\bar\chi_{ij}&=&\tilde\chi_{ij},\label{outsidechi}\\
\bar\Delta_{ij}&=&\tilde\Delta_{ij}\label{outsideDelta}.
\en
Recalling that $\bar\chi_{ij}$ and $\bar\Delta_{ij}$ are interpreted as the
hopping and pairing amplitudes of physical electrons, we see that
{\it the region 
outside the SF core and inside the gauge field penetration depth,
$\ell_c \alt r \alt \lambda_a$ around the vortex,
 is characterized by
the staggered  modulation of the hopping and pairing amplitudes}.
Note that the amplitude of $\bar\chi_{ij}$ and
$\bar\Delta_{ij}$ are modulated in a correlated way 
according to Eq.~(\ref{modchiDelta}) 
to preserve
\be
\tilde\chi_{ij}^2+\tilde\Delta_{ij}^2={\rm constant}.\label{corr}
\en
In Fig.~3, we depict the situation 
given by Eqs.~(\ref{modchiDelta}), 
(\ref{outsidechi}),  (\ref{outsideDelta}),
and (\ref{corr}).
%%%%%%%%%%%%%%%%%%%%
%\noindent Fig.3
\begin{figure}
%\epsfxsize=2.0in
%\centerline{\epsffile{Fig03.eps}}
\smallskip
\caption{
Geometric relation of $\tilde\chi_{ij}$
and $\tilde\Delta_{ij}$.
The angle $\Phi_{ij}$ modulates around $\Phi_0$
in a staggered manner [see Eq.~(\ref{Phiij})].
}
\end{figure}
%                 %
%%%%%%%%%%%%%%%%%%%
%%
$\bar U_{ij}^d$ breaks
 the translational symmetry,
but does not break
the time reversal symmetry with  respect to the local bonds.
Therefore, $\bar U_{ij}^d$ does not cause local fermion
current  on the bonds [
of course, even in this case,
the external magnetic field breaks time reversal symmetry
and causes globally circulating supercurrent given by Eq.~(\ref{gcu})].
What is staggering in this  region
is not the local current but the local density on the 
bonds.

The temporal component of the gauge field,
${\bb{a}}_{0i}^d$, is determined locally by the LBC.
In the uniform case, the saddle point is purely imaginary. 
There, we can regard the LBC as almost uniform in the SC state
outside the vortex core. Therefore it may be legitimate to
assume  ${\bb{a}}_{0i}^d$ to be uniform and parallel to the LBC (pointing
toward the north pole)
in this region [this assumption is reliable as far as
deviation of $\theta_i$ from $\sim\pi/2$ is small].
From now on, we set 
\be
i{\bb{a}}_{0i}^d=(0,0,a_0),\label{optadsc}
\en
where $a_0$ has an order of $x\tilde J$.

For the purpose of seeing physical situation, 
we assume the SF core size to be $\ell_c=3$ and 
a simple distribution of the angle $\theta$ as indicated in Fig.~4(a).
In Figs.~4(b) and 4(c), respectively,
 we show the corresponding spatial variation
of  $\bar\chi_{ij}$ and $\bar\Delta_{ij}$
on the link connecting $(i_x,0)$ and $(i_x,1)$ with lattice unit.
We also assumed the gauge field penetration depth to be $\lambda_a=10$.
Now $\bar\Delta_{ij}$  just represents pairing amplitude of the physical electron
and vanishes at the vortex center as it should do.
As we go away from the core, 
$\bar\chi_{ij}$ and $\bar\Delta_{ij}$ acquire staggered modulation with 
the amplitude becoming smaller, because
the superfluid velocity $\bb{v}(\r)$, which is
responsible for appearance of the staggered modulation,
becomes smaller.
In fact, the staggered modulation of $\bar\chi_{ij}$
is just an order of a few percent,
while that of $\bar\Delta_{ij}$ is rather large.
However, as we shall see shortly,
{\it period doubling} caused by this modulation
gives rise to visible effects in local density-of-states (LDOS) 
{\it outside the core.} We should also remark that both
$\bar\chi_{ij}$ and $\bar\Delta_{ij}$ contribute to the LDOS.
 The problem now reduces  to
the more familiar U(1) mean field theory, but with $\chi_{ij}$ and
$\Delta_{ij}$
which vary in space. 
This is precisely the problem treated by Han, Wang, 
and
Lee\cite{Han-Wang-Lee00,Wang-Han-Lee01}and it is gratifying that they found
numerically the  staggered current around the vortex core as
proposed in the SU(2) vortex model.\cite{Lee-Wen00}
%%%%%%%%%%%%%%%%%%%%
%\noindent Fig.4
\begin{figure}
%\epsfxsize=3.3in
%\centerline{\epsffile{Fig04.eps}}
\smallskip
\caption{
 A simple distribution of the $\theta$ indicated in (a)
  leads to  
spatial variation of (b) $\bar\chi_{ij}$ and (c) $\bar\Delta_{ij}$
on the link connecting $(i_x,0)$ and $(i_x,1)$ as
indicated in the inset. The gauge field penetration depth is
 assumed to be $\lambda_a=10$.
}
\end{figure}
%                 %
%%%%%%%%%%%%%%%%%%%

In Fig.~5, we schematically show the modulation pattern of
$\bar\chi_{ij}$ outside the core. The staggered modulation  becomes
most conspicuous when scanned along the straight line
$i_x=1/2$ or $i_y=1/2$, provided that the vortex center sits at $(1/2,1/2)$,
 because on these bonds
 the circulating ${\bv}({\r})$-field becomes parallel to the bond 
directions.
%As we have already seen in Fig.~4,
%the modulation amplitudes of $\chi_{ij}$ and $\Delta_{ij}$ are 
%small as compared with the original magnitude of $\chi$ and $\Delta$.
%We shall see  in the next section, however,
%the period doubling gives rise to very specific structure
%in the LDOS outside the core.
Apparently, the bond modulation pattern  reminds us of
the spin-Peierls states.
However, this is not the case,
 since the MF expectation value of spin-exchange energy
on the bonds
is given by
$\langle{\bbox{S}}_i\cdot {\bbox{S}}_j\rangle=
-\tilde J (\bar\chi_{ij}^2 + \bar\Delta_{ij}^2)=\rm{const.}$
and therefore the spin-Peierls order parameter becomes
$\langle{\bbox{S}}_i\cdot {\bbox{S}}_{i+\hat{\bf{e}}_\mu}-
{\bbox{S}}_i\cdot {\bbox{S}}_{i-\hat{\bf{e}}_\mu}\rangle=0$
with $\hat{\bf{e}}_\mu$ being a unit vector connecting the neighboring sites.
%%%%%%%%%%%%%%%%%%%%
%\noindent Fig.5
\begin{figure}
%\epsfxsize=3.2in
%\centerline{\epsffile{Fig05.eps}}
\smallskip
\caption{
Schematic drawing of
the amplitude modulation pattern of the hopping  parameter
$\tilde\chi_{ij}$ outside the SF core.
Solid and dotted bonds indicate
 enhanced and reduced amplitudes, respectively, where
thickness of the bonds qualitatively represents magnitude of the 
modulation.
Circulation of the 
the fermion \lq\lq superfluid velocity\rq\rq,
$\bv({r})$, 
associated with the internal gauge field $\bb{a}^3$
 is  indicated by the arrows.
 The staggered modulation  becomes
most conspicuous when scanned along the lines 
$i_x=1/2$ or $i_y=1/2$, 
provided that the vortex center sits at $(1/2,1/2)$.
 Note 
that the boundary of the SF core region,
inside which the staggered orbital currents flow,
should not be taken literally. 
In reality,
there is a crossover region around
$r\sim \ell_c$ where the staggered
current and the staggered amplitude modulation coexists.
}
\end{figure}
%                 %
%%%%%%%%%%%%%%%%%%%

As we approach the core from the outside,
the ${\bbox I}$-vector in the SF gauge  gradually rises 
off 
from the equatorial plane [see Fig.1~(a)].
This may  give rise to a
 crossover region characterized by
 coexistence of the amplitude and phase modulation,
 where  $\theta$-dependence of $\bar U_{ij}^d$
 becomes significant.
 It is expected that the staggered current begins to appear
 around $r\sim \ell_c$ and its strength becomes stronger as we approach the
 immediate center of the vortex.
 We give a schematic drawing of this circumstance in Fig.~5.
To study the effects of $\theta$-dependent $\bar U_{ij}^d$
is, however,  beyond the scope of the present paper
and  we concentrate on the region 
$\ell_c\alt r\alt \lambda_a$.
We should also remark that 
when the angle $\theta$ deviates from $\theta_i=\theta_j=\pi/2$
as we approach the core,
the direction $\bb{a}_{0i}^d$ begins to
slightly deviate from the north pole, since
$\bb{a}_{0i}^d$ is no longer
parallel to the $\bb{I}$-vector due to small anisotropy.  
In the next section, we shall  compute the LDOS in the SC state outside
the core
by setting $\theta_i=\theta_j=\pi/2$.
Then, $\bb{a}_{0i}^d$ is given by Eq.~(\ref{optadsc}) and exactly 
parallel to the $\bb{I}$-vector pointing toward the north pole.
We expect our results to be qualitatively valid even for
$r \sim \ell_c$ as long as we avoid the inside of the core.

%%%
%%%
%%%
%%%

\section{LDOS outside the core}
As we saw in the previous section,
the staggered modulation of the hopping amplitude $\bar\chi_{ij}$
and pairing amplitude $\bar\Delta_{ij}$  becomes predominant
over the region $\ell_c\alt r\alt \lambda_a$.
The presence of staggered modulation
 suggests that this
 may be the best place
 to look for unit cell doubling effect.
In this section, we consider the LDOS in this region.

\subsection{Formulation of LDOS}

The local density of states at an arbitrary point $\r$ on lattice
is defined by
\be
N(\r,\omega)=-{2\over\pi}{\rm Im}{\cal G}^{\rm phys}(\r, \r; i\omega)
\mid_{i\omega=\omega+i\delta}\label{originalLDOSformula}
,
\en
where the propagator for the physical electron is introduced 
by
$
{\cal G}^{\rm phys}(\r,\r',i\omega)=-\int_0^\beta d\tau e^{i\omega\tau}
< T_\tau c_\sigma(\r,\tau)c_\sigma^\dagger(\r')>.
$
To model the tunneling current we assume that the electron tunnel from the
tip located at
 $\bbox r$ to a linear combination of Wannier orbitals centered at lattice
sites, 
i.e., 
the physical electron operator at $\r$,
$c_\sigma(\r)$, is related to $c_{i\sigma}$ as
\be
c_\sigma(\r)=\sum_{i}\alpha_i(\r)
e^{-i{e\over c}\int_{\r_i}^{\r}d\r' \cdot \bb{A}(\r')} c_{i\sigma}
,
\label{wannier}
\en
where  the EM gauge potential $\bb{A}$
gives rise to the EM Peierls phase.
The envelope function $\alpha_i(\r)$ may be simulated by
$\alpha_i(\r)=e^{-|\r-\r_i|/\xi}$ in the bond direction
(the Cu-O-Cu bond). 
The length scale $\xi$ can reasonably be 
 set equal to $\xi=1/2$ with lattice scale, corresponding to
 the Cu-O separation.
 Since the effects of the EM gauge fields are
 negligibly small in strength as compared with the internal gauge
potential,
from now on, we ignore the EM Peierls phase and examine the effects of
the staggered hopping and pairing   amplitudes
on the LDOS.
Noting Eq.~(\ref{physhole}) in the $d$-wave gauge,
Eq.~(\ref{originalLDOSformula}) is  written as
\be
N({\bbox r},\omega)=-{x\over \pi}{\rm Im}
\sum_{i,j}\alpha_i({\bbox r})\alpha_j({\bbox r})
[{\cal G}_{ij}^{\rm F}(\ii\omega)]_{11}
|_{\ii\omega=\omega+\ii\delta}.\label{startingformula}
\en
The subscript 11 means 11-component of the
lattice fermion propagator of a $2\times 2$ matrix form,
$
{\cal G}_{ij}^{\rm F}(\tau)=
-
\langle T_{\tau} \psi_{i\sigma}(\tau)\psi_{j\sigma}^\dagger\rangle.
$

We here give an intuitive demonstration  that
the LDOS  exhibits conspicuous staggered pattern
{\it only when measured on the bonds}.
More quantitative discussion will be given in the following subsections.
For example,
we pick up the sites $1,2,...,6$ indicated in Fig.~6
and consider the midpoints on the bond, B$_1$, B$_2$,
and the plaquette centers, C$_1$, C$_2$.
The LDOS  at C$_1$ and C$_2$ come from
 $
\sum_{i,j=1,2,4,5}
{\cal G}_{ij}^{\rm F}$,
and  $
\sum_{i,j=2,3,5,6}
{\cal G}_{ij}^{\rm F}$,
respectively. We see, however,
${\cal G}_{12}^{\rm F}\sim{\cal G}_{56}^{\rm F}$
because the bonds 12 and 56 are
almost equivalent except 
 the effects of 
 negligibly  small dependence of the 
$\bv$-field on the spatial position $\r$ over
the lattice scales.
 Similarly,
${\cal G}_{45}^{\rm F}\sim{\cal G}_{23}^{\rm F}$ and
${\cal G}_{14}^{\rm F}\sim{\cal G}_{36}^{\rm F}$.
Therefore,  $N(\r_{{\rm C}_1},\omega)\sim N(\r_{{\rm C}_2},\omega)$.
Similarly, the LDOS
at the lattice sites is almost uniform.
On the other hand, the LDOS
at B$_1$ and B$_2$ come from
 $
\sum_{i,j=1,2}
{\cal G}_{ij}^{\rm F}$,
and
$
\sum_{i,j=2,3}
{\cal G}_{ij}^{\rm F}$,
respectively.  Here, ${\cal G}_{12}^{\rm F}$ and
${\cal G}_{23}^{\rm F}$ are 
clearly inequivalent 
because
they connect the bonds with alternating hopping-pairing amplitudes.
%%%%%%%%%%%%%%%%%%%%
%Fig.6
\begin{figure}
%\epsfxsize=2in
%\centerline{\epsffile{Fig06.eps}}
\smallskip
\caption{
Points on lattice where we consider the LDOS.
We have four symmetrically distinct points:
the plaquette-center ($\times$), site-top ($\bullet$), and
bond-center ($\circ$).
The site-top and bond-center points correspond to
the Cu and O sites, respectively.}
%\label{short-fig1}
\end{figure}
%                 %
%%%%%%%%%%%%%%%%%%%
To compute
 ${\cal G}_{ij}^{\rm F}(\ii\omega)$ in the SC state outside the core,
we  shall use  the following two approaches which may be complementary to
each other:
(I) {\it a perturbative analysis using the gradient expansion }, and
(II)  {\it an exact diagonalization using the \lq\lq uniform $\bv$\rq\rq
approximation }.
In the former approach,  we can take  account of the circulating
configuration of the $\bv({\bbox r})$-field, while
in the latter approach, instead, we can obtain a nonperturbative aspect
of the problem.

\subsection{Perturbative analysis using the gradient expansion}

First,
we expand (\ref{modchiDelta}) with respect to $v_{ij}$ up to
the first order
 as
\be
\bar\chi_{ij}&\sim&\chi_0-
{(-1)^{i_x+j_y}}
\Delta_0 v_{ij}
,\\
\bar\Delta_{ij}&\sim&\Delta_0+
{(-1)^{i_x+j_y}}
\chi_0 v_{ij},
\en
which give
$
\bar U_{ij}^d
=U_{ij}^{d}
+\delta U_{ij}^d
$
with
\be
\delta U_{ij}=
{(-1)^{i_x+j_y}}
[\Delta_0\tau^3
+(-1)^{i_y+j_y}\chi_0\tau^1]v_{ij}.
\en
Then, we treat the term
\be
\delta H^{\rm F}
={\tilde J\over 2}\sum_{<i,j>}
\psi_i^\dagger\delta U_{ij}\psi_i,\label{perturbation}
\en
as perturbation with respect to
$
H_0^{\rm F}
$
where ${\bb{a}}_{0i}$  is given by Eq~(\ref{optadsc}).
The free propagation  is governed by $U_{ij}^d$
and the corresponding propagator becomes
\be
{\cal G}^{\rm F}_0({\bbox k},\ii\omega)
=
{U_{\bbox k}\over i\omega-E_{\bbox k}}
+{V_{\bbox k}\over i\omega+E_{\bbox k}},\label{freepropagator}
\en
where the generalized coherence factors are introduced by
$
U_{\bbox k}={1\over2}
[1+{(\gamma_{\k}\tau^3+\eta_{\k}\tau^1)/ E_{\bbox k}}],
$
and
$
V_{\bbox k}=
{1\over2}
[1-{(\gamma_{\k}\tau^3+\eta_{\k}\tau^1)/E_{\bbox k}}].
$
The one-particle spectrum is given by
\be
E_{\bbox k}=\sqrt{\gamma_k^2+\eta_k^2},\label{gamma}
\en
with
$
\gamma_{\bbox k}=-\tilde J\chi_0[
\cos k_x+\cos k_y
+\tilde t_2 \cos k_x\cos k_y
+\tilde t_3(\cos 2 k_x+\cos 2 k_y)]+a_0,
$
and
$
\eta_{\bbox k}=+\tilde J\Delta_0(\cos k_x-\cos k_y).
$
We took account of the second and third nearest neighbor hopping of the
fermions to reproduce the
real band structure.
In general, the $d$-wave nodes shift from $(\pm\pi/2,\pm\pi/2)$.
In the case of $t_2=t_3=0$, the nodes are located at
$
(\pm\cos^{-1}[a_0/ 2\tilde J\chi_0],\pm\cos^{-1}
[a_0/ 2\tilde J\chi_0]).
$ 
For $\tilde t_2\neq 0$ and $\tilde t_3\neq 0$,
the nodes are located at
$
(\pm\cos^{-1}[f(\tilde t_2,\tilde t_3,a_0)],\pm\cos^{-1}
[f(\tilde t_2,\tilde t_3,a_0)]), 
$
where
\be
f(\tilde t_2,\tilde t_3,a_0)={-1+
\sqrt{1+(4\tilde t_3+\tilde t_2)(2\tilde t_3+a_0/\tilde J\chi_0)}\over
4\tilde t_3+\tilde t_2}
.
\en
We see that as far as $\tilde t_3\neq 0$
the nodes shift from $(\pm\pi/2,\pm\pi/2)$ even if $a_0=0$.
Furthermore we note that
location of the nodes is independent
of the gap magnitude $\Delta$.

Since the perturbation term causes period doubling, it is convenient to
introduce
the fermion operators on
two sublattices:
\be
\psi_{i}={1\over\sqrt{2}}
\sum_{\k\in\rm RZ}e^{\ii\k\cdot\r_i}
\left(\psi_{\k}\pm\psi_{\k+\bb{Q}}\right),\label{wfonsublattice}
\en
where $\bb{Q}=(\pi,\pi)$
and $\k\in \rm RZ$ means $\k$ runs over the reduced Brillouin zone
$|k_x|+|k_y|\leq\pi$. We dropped the spin indices.
The $+$ and $-$ signs are for the cases where
$\bb{i}$ belongs to the A [$\r_i=(i_x,i_y)=({\rm even},{\rm even})$
or $({\rm odd},{\rm odd})$]
and B [$(i_x,i_y)=({\rm even},{\rm odd})$
or $({\rm odd},{\rm even})$] sublattice sites, respectively.
Then, as derived in appendix C, 
the perturbation term is written in momentum space as
\be
\delta H^{\rm F}
=-
\sum_{{\bbox k}\in \rm RZ}
\sum_{{\bbox q},\sigma}[
\psi_{{\bbox k}+{{\bbox q}\over2}+{\bbox Q}\sigma}^{\dagger}
{\cal C}_{\bbox k}({\bbox q})\psi_{{\bbox k}-{{\bbox q}\over2}\sigma}
+{\rm H.c.}],\label{conpert}
\en
where
$
{\cal C}_{\bbox k}({\bbox q})
=\Delta_0  {\cal C}^+_{\bbox k}({\bbox q})\tau^3+\chi_0 {\cal C}^-_{\bbox k}({\bbox 
q})\tau^1,
$
with
\be
{\cal C}^\pm_{\bbox k}({\bbox q})
=
\pi \tilde J
 {1\over |{\bbox q}|^2}
{\lambda_a |{\bbox q}|\over 1+\lambda_a |{\bbox q}|}
\left[
q_y \sin k_x\pm q_x \sin k_y\right].
\en
The momentum transfer $\q$ should be small because we retained only slowly
varying
$\bv$-field.
The perturbation processes cause the unit cell doubling and
scatter
the electron with $\k$ in the reduced zone to $\k+\bigQ$ in the second 
zone,  and consequently the mirror image
of the reduced zone is formed in the second zone,
as indicated in Fig.~7(a).
%%%%%%%%%%%%%%%%%%%%
%Fig.7 %
\begin{figure}
%\epsfxsize=3in
%\centerline{\epsffile{Fig07.eps}}
\smallskip
\caption{
(a)
The perturbation processes given by Eq.~(\ref{perturbation})
 connect
the electron with $\k$ in the reduced zone [inner square]
to $\k+\bigQ$ in the second zone [shaded region], 
and consequently the mirror image
of the reduced zone is formed in the second zone.
The $d$-wave nodes inside the reduced zone
and their mirror images are also indicated.
(b) The scattering processes
along the $(0,0)\to(\pi,\pi)$ direction
whose matrix elements
give the coherence factors
$
L_{\k}^\pm
$ and
$
N_{\k}^\pm.
$
At the energy $\omega^\ast$,  the Dirac cones around the $d$-wave nodes 
touch the reduced zone boundary and
the resonance occurs.
(c) The level crossing at the reduced zone boundary
would be lifted and eventually the period doubling would cause
gap opening 
if we would go beyond the perturbative scheme.
Note that situations in 
(b) and (c)  correspond to
the case of a simple
band structure without the next 
($\tilde t_2$) and second nearest ($\tilde t_3$) fermion hopping.
}
%\label{short-fig1}
\end{figure}
%                 %
%%%%%%%%%%%%%%%%%%%

Now, we consider the four distinct points on lattice
indicated in Fig.~6:
(a) the center of the plaquette (plaquette-center),
(b) the top of the sites (site-top), and
(c) the center of the bonds (bond-center).
The site-top and the bond-center points correspond to
the Cu and the O sites, respectively, on the CuO$_2$ plane.
All the detail of derivation of the LDOS is left to
 appendix C. 
In any case,
the LDOS is written in a form
\be
N({\bbox r}, \omega)/
x \alpha^2
=
\bar N_0(\omega)\pm\delta \bar N({\bbox r}, \omega),\label{LDOSformula}
\en
where $+$ and $-$ signs alternate from plaquette to plaquette,
site to site, or bond to bond for the cases
(a), (b) and (c), respectively,
and
$\alpha$ represents magnitude of
 the envelope function from
the nearest site.
The uniform  counterpart is given in a form
\be
\bar N_0(\omega)
=
-{1\over\pi}{\rm Im}
\sum_{\k} M_0(\k)
[{\cal G}^{\rm F}(\k,\k,\ii\omega)]_{11}|_{\ii\omega=\omega+\ii\delta},
\label{formulauniform}
\en
where
we introduced the generalized propagator
$
{\cal G}^{\rm F}(\k,\k',\ii\omega)
=
\sum_{{\r}_i,{\r}_j}
e^{\ii(\k\cdot\r_i-\k'\cdot \r_j)}
{\cal G}_{ij}^{\rm F}(\ii\omega).
$
The matrix elements $M_0(\k)$ distinguishes different
symmetries
associated with each point
and are given by
\be
&&M_0^{\rm plaquette}
(\k)=\cos^2{k_x\over2}\cos^2{k_y\over2},\label{M0a}\\
&&M_0^{\rm site}(\k)=1,\\
&&M_0^{\rm bond}(\k)=\cos^2{k_x\over2}\label{M0c},
\en
 for the cases (a), (b), and (c),
 respectively.
The perturbation processes do not affect
the uniform counterpart within the Born approximation, and 
thus, in Eq.~(\ref{formulauniform}),
we obtain
\be
&&{\rm Im}[{\cal G}^{\rm F}
(\k,\k,\ii\omega)]_{11}|_{\ii\omega=\omega+\ii\delta}
={\rm Im}
[{\cal G}_0^{\rm F}(\k,\ii\omega)]_{11}|_{\ii\omega=\omega+\ii\delta}
\\
&=&
{\pi\over 2}
\left(1+{\gamma_k\over E_{\k}}\right)
\delta(\omega-E_{\k})
+{\pi\over 2}
\left(1-{\gamma_k\over E_{\k}}\right)
\delta(\omega+E_{\k}),\non
\en
which just reproduces the
LDOS profile in the uniform $d$-wave SC state except overall
reduction due to the matrix element $M_0(\k)$.

The staggered counterpart is given in a form
\be
\delta \bar N(\r,\omega)
=
-{1\over\pi}{\rm Im}
\sum_{\q\sim\,\,\rm small}
\sum_{\k\in\rm RZ} M(\k,\q;\r)
[
{\cal G}^{\rm F}
(\k+{\q\over 2}+\bb{Q},\k-{\q\over 2},\ii\omega)
+
{\cal G}^{\rm F}
(\k-{\q\over 2},\k+{\q\over 2}+\bigQ,\ii\omega)
]_{11}
\mid_{\ii\omega=\omega+\ii\delta},\label{formulastagg}
%%%%%%%%%%%%%%%%%
\en
The matrix elements $M(\k,\q;\r)$ 
associated with each point
are given by
\be
&&M^{\rm plaquette}(\k,\q;\r)=
\cos(\q\cdot \r)
%\non\\&&\,\,\,\,\,\,\times
\sin{k_x+{q_x\over 2}\over2}\sin{k_y+{q_y\over 2}\over2}\cos{k_x-{q_x\over
2}\over2}\cos{k_y-{q_y\over 2}\over2},\label{casea}\\
&&M^{\rm site}(\k,\q;\r)=
\cos(\q\cdot \r)\label{caseb},\\
&&M^{\rm bond}(\k,\q;\r)=
\sin(\q\cdot \r)
\sin{k_x+{q_x\over 2}\over2}
\cos{k_x-{q_x\over 2}\over2}\label{casec}
,
\en
where $\r$ denotes the plaquette-center, site-top, and bond-center points,
 respectively.
Now we need to compute
$
{\cal G}^{\rm F}
(\k+{\q\over 2}+\bb{Q},\k-{\q\over 2},\ii\omega)
$
and
$
{\cal G}^{\rm F}
(\k-{\q\over 2},\k+{\q\over 2}+\bb{Q},\ii\omega).
$
The detail of computation 
is presented in appendix C. 
We obtain
\be
&&
[
{\cal G}^{\rm F}
(\k+{\q\over 2}+\bb{Q},\k-{\q\over 2},\ii\omega)
%\non\\&&\,\,\,\,\,\,\,\,\,\,\,\,\,\,\,
+{\cal G}^{\rm F}
(\k-{\q\over 2},\k+{\q\over 2}+\bigQ,\ii\omega)
]_{11}
\mid_{\ii\omega=\omega+\ii\delta}
\non
\\
&=&
%%%
-{\pi \over 2}
\delta(\omega,+E_{\k+\q/2+\bigQ},+E_{\k-\q/2})
%\non\\&&\,\,\,\,\,\,\,\,\,\,\,\,\,\,\,\,\,\,\,\,\,\,\,\,\,\,\,\times
\left[
\Delta_0 {\cal C}_{\k}^+(\q)
L^{\Delta+}_{\bbox k}
+\chi_0 {\cal C}_{\k}^-(\q)
L^{\chi+}_{\bbox k}
\right]
\non\\
&+&
%%%
{\pi \over 2}
\delta(\omega,-E_{\k+\q/2+\bigQ},-E_{\k-\q/2})
%\non\\&&\,\,\,\,\,\,\,\,\,\,\,\,\,\,\,\,\,\,\,\,\,\,\,\,\,\,\,\times
\left[
\Delta_0 {\cal C}_{\k}^+(\q)
L^{\Delta-}_{\bbox k}
+\chi_0 {\cal C}_{\k}^-(\q)
L^{\chi-}_{\bbox k}
\right]
\non\\
&-&
%%%
{\pi \over 2}
\delta(\omega,+E_{\k+\q/2+\bigQ},-E_{\k-\q/2})
%\non\\&&\,\,\,\,\,\,\,\,\,\,\,\,\,\,\,\,\,\,\,\,\,\,\,\,\,\,\,\times
\left[
\Delta_0 {\cal C}_{\k}^+(\q)
N^{\Delta+}_{\bbox k}
+\chi_0 {\cal C}_{\k}^-(\q)
N^{\chi+}_{\bbox k}
\right]
\non\\
&+&
%%%
{\pi \over 2}
\delta(\omega,-E_{\k+\q/2+\bigQ},+E_{\k-\q/2})
%\non\\&&\,\,\,\,\,\,\,\,\,\,\,\,\,\,\,\,\,\,\,\,\,\,\,\,\,\,\,\times
\left[
\Delta_0 {\cal C}_{\k}^+(\q)
N^{\Delta-}_{\bbox k}
+\chi_0 {\cal C}_{\k}^-(\q)
N^{\chi-}_{\bbox k}
\right],\label{perturbationresult}
\en
where
$\delta(\omega,x,y)\equiv[\delta(\omega-x)-\delta(\omega-y)]/(x-y).$
The first equality in   Eq.~(\ref{perturbationresult}) just represents the
inversion and
time reversal symmetries of the propagator:
$
{\cal G}^{\rm F}
(\k+{\q\over2}+\bigQ,\k-{\q\over2},\ii\omega)=
{\cal G}^{\rm F}
(\k-{\q\over2},\k+{\q\over2}+\bigQ,\ii\omega).
$
The coherence factors are given by
\be
L^{\Delta\pm}_{\bbox k}
&=&
1+{\gamma_+}{\gamma_-}-{\eta_+}{\eta_-}\pm{\gamma_+}\pm{\gamma_-},\non\\
L^{\chi\pm}_{\bbox k}&=&
\pm{\eta_+}\pm{\eta_-}+
{\gamma_+}{\eta_-}+{\eta_+}{\gamma_-},\non\\
N^{\Delta\pm}_{\bbox k}&=&
1-{\gamma_+}{\gamma_-}+{\eta_+}{\eta_-}\pm{\gamma_+}\mp{\gamma_-},\non\\
N^{\chi\pm}_{\bbox k}&=&
\pm{\eta_+}\mp{\eta_-}-{\gamma_+}{\eta_-}-{\eta_+}{\gamma_-},
\en
where
$
\gamma_+={\gamma_{{\bbox k}+\q/2+{\bbox Q}}
/ E_{{\bbox k}+\q/2+{\bbox Q}}},
$
$
\gamma_-={\gamma_{\k-\q/2}/ E_{\k-\q/2}},
$
$
\eta_+={\eta_{{\bbox k}+\q/2+{\bbox Q}}/ E_{{\bbox k}+\q/2+{\bbox Q}}},
$
and
$
\eta_-={\eta_{\k-\q/2}/ E_{\k-\q/2}}.
$

To proceed with further analytical computation,
we note that main contribution of $\k$
 integral comes from regions near the nodes in the vicinity of
$(\pm\pi/2,\pm\pi/2)$, while $\q$ is small.
Thus, it is legitimate to ignore $\q$ with respect to $\k$
in $M_0(\k)$, $M(\k,\q;\r)$, and
${\cal G}^{\rm F}
(\k+{\q\over2}+\bigQ,\k-{\q\over2},\ii\omega)$, while we must retain
$\q$ in
$
{\cal C}_{\k}^\pm(\q)
$.
This approximation amounts to
ignoring  $\r$-dependence of $\bv(\r)$-field  over the lattice scales, and
retaining only fermion fluctuations. On the other hand,
 retaining 
 $\q$-dependence of ${\cal C}_{\k}^\pm(\q)$ amounts to
 taking account of long distance decay of the $\bv$-field.
Under this approximation,
 Eqs.~(\ref{casea}), (\ref{caseb}), and (\ref{casec}) are simply
 reduced to
\be
M^{\rm plaquette}(\k,\q;\r)&=&{1\over 4}
\cos(\q\cdot \r)
\sin{k_x}\sin{k_y},\\
M^{\rm site}(\k,\q;\r)&=&
\cos(\q\cdot \r),\\
M^{\rm bond}(\k,\q;\r)&=&{1\over 2}
\sin(\q\cdot \r)
\sin{k_x}.\label{simpbondmatel}
\en
By noting the antisymmetry relation
${\cal C}_{\k}^\pm(\q)=-{\cal C}_{\k}^\pm(-\q)$,
we immediately see that
$\delta\bar N(\r,\omega)$ vanishes
at the plaquette-center and site-top points 
while it remains finite  at the bond-center points.
Thus, we confirm that {\it the staggered counterpart of the LDOS
appears only when measured on the bonds}. 
Even in the cases of the plaquette-center and the site-top,
$\delta\bar N(\r,\omega)$  becomes  finite
if we retain $\q$ with respect to
$\k$, i.e., take account of negligibly  small dependence of the 
$\bv$-field on the spatial position $\r$ over
the lattice scales.
However, this effect is still invisibly small as compared with 
the case of the bond-center.
This result is fully consistent
with an intuitive discussion given in Sec. IV A.

From now on, we concentrate on the bond-center points:
 the midpoint of the bond connecting
$\bb{i}$ and $\bb{i}+\hat{\bf{e}}_\mu$ where
$\mu=x$ or $y$.
Taking account of the envelope function,
the magnitude of the LDOS 
may be reduced by a factor $e^{-2}\sim 0.1$
as compared with the uniform counterpart of the LDOS at the
site-top.
%%%%%%%%%%%%%%%%%%%%%
Using Eqs.~(\ref{perturbationresult}) and (\ref{simpbondmatel}),
the $\q$-integration in Eq.~(\ref{formulastagg})
can be performed
to yield
\be
&&\delta \bar N({\bbox  r},\omega)
=
{(-1)^{i_x+i_y}\over 4}
v_{\mu}(\r)
\sum_{{\bbox k}\in\rm RZ}
\sin^2 k_{\mu}\\
&&
\left[
L_{\bbox k}^+ \delta(\omega;E_{\bbox k},E_{{\bbox k}+{\bbox Q}}
)
\right.
+
%%%
L_{\bbox k}^-
\delta(\omega;-E_{\bbox k},-E_{{\bbox k}+{\bbox Q}})
\non\\
&&+
%%%
N_{\bbox k}^+
\delta(\omega;E_{\bbox k},-E_{{\bbox k}+{\bbox Q}})
+\left.
%%%
N_{\bbox k}^-\delta(\omega;-E_{\bbox k},E_{{\bbox k}+{\bbox Q}})
\right]\non,
%%%%%%%%%%%%%%%%%
\en
where the coherence factors
$
L_{\bbox k}^\pm
=
\Delta_0 L_{\bbox k}^{\Delta\pm}
+
\chi_0 L_{\bbox k}^{\chi\pm},
$ and
$
N_{\bbox k}^\pm
=
\Delta_0 N_{\bbox k}^{\Delta\pm}
+
\chi_0 N_{\bbox k}^{\chi\pm}
$
represent the matrix element associated with
the scattering processes  indicated in Fig.~7(b).
As has already been mentioned,
the best paths to detect 
the staggered modulation of the  LDOS are
the lines $i_x=1/2$ or $i_y=1/2$ provided that the vortex center
sits at $(1/2,1/2)$,
because in this case we can go through
the bonds whose directions ${\hat{\bf{e}}}_\mu$
are parallel to the circulating ${\bv}({\r})$-field [see Fig.~5].

\subsubsection{The case of $t_2=t_3=0$}
First, we consider a toy band structure
with  $t_2=t_3=0$ in
Eq.~(\ref{gamma}), because
this simple case  provides us with
a clear view on  the period doubling effects.
 In Fig.~8(a), we show the profile of
 $N({\bbox  r},\omega)/x \alpha^2$
 at the four bond-center
 points,  A$(i_x,1/2)$, B$(i_x+1/2,0)$, 
 C$(i_x+1,1/2)$, and
D$(i_x+1/2,1)$ with $i_x=5$ [see the inset of Fig.~8(a)].
From now on, 
 we fix the parameters
 $a_0=0.05 \chi_0 \tilde J$,
 $\Delta_0/\chi_0=0.2$, and assume the gauge field penetration depth to be
$\lambda_a=10$.
 This choice of $a_0$ and $\Delta_0$ is reasonable in
the underdoped regime.\cite{Wen-Lee96}
%Numerical integration was performed by dividing the Brillouin zone into
%$320\times 320$ meshes.
 Note that at B and D,
 $\delta \bar N({\bbox  r},\omega)$
 almost vanishes and the LDOS is just given by
 $\bar N_0(\omega)$,
because  $\bv(\r)$ becomes almost perpendicular to these bond directions.
 The modulation pattern at the other points can  be  read off from Fig.~5.

%%%%%%%%%%%%%%%%%%%%
%Fig.8 %
\begin{figure}
%\epsfxsize=3in
%\centerline{\epsffile{Fig08.eps}}
\smallskip
\caption{
(a) LDOS profile in the case of $t_2=t_3=0$, obtained by the perturbative 
analysis at the points
A, B, C, and D indicated in the inset.
The LDOS at B and D are just $\bar N_0(\omega)$.
The peaks at $\tilde\omega\equiv\omega/\chi_0\tilde J=\pm 0.38$ 
are associated with the $d$-wave
superconducting gap.
The additional peaks at $\tilde\omega=\pm 0.41$
are associated with the van-Hove singularity located at $(0,\pm\pi)$ and $(\pm\pi,0)$.
The staggered   structure around $\tilde\omega=0.05$ comes from
 resonant scattering between the fermions with
${\bbox k}$ and ${\bbox k}+{\bbox Q}$, caused by the period doubling.
(b) The one-particle energy contour
around the $d$-wave node.
(c) The energy contours $E_{\k}=\omega$ and
$E_{{\k}+{\bigQ}}=\omega$ touch at $\tilde\omega=\pm 0.05$.
}
%\label{short-fig1}
\end{figure}
%                 %
%%%%%%%%%%%%%%%%%%%

We see that
 inside the overall V-shaped profile with the sharp peaks at
$\tilde\omega\equiv\omega/\chi_0 \tilde J=\pm 0.38$
associated with the $d$-wave
superconducting
gap, there appears additional peak and dip
structure at site C and A, respectively,
 around $\tilde\omega=+0.05$.
 From now on, we refer to this structure as 
 \lq\lq staggered peak-dip (SPD)\rq\rq structure, since
 the peak and dip alternate from bond to bond in a staggered manner.
The additional peaks at $\tilde\omega=\pm 0.41$
come from the van-Hove singularity located at $(0,\pm\pi)$ and $(\pm\pi,0)$
points.
%, which is intrinsic to the normal state dispersion $\gamma_{\bf{k}}$.
The low energy dispersion gives
elliptic contours around the $d$-wave node as indicated in Fig.~8(b)
 which touch the reduced zone boundary
 at $(\pi/2,\pi/2)$
as the energy increases.
The specific  structure around $\tilde\omega=0.05$
 comes from resonant scattering between the fermions with
${\bbox k}$ and ${\bbox k}+{\bbox Q}$.
As $\omega$ increases from  zero, the
 energy contours
$E_{{\bbox k}}=\omega$ and $E_{{\bbox k}+{\bbox Q}}=\omega$ touch
at $(\pi/2,\pi/2)$ on the reduced zone boundary at $\tilde\omega=\pm 0.05$
as indicated in Fig.~8(c) [see also Fig.~7(b)] and resonance occurs.
We note that
the modulated structure inside the V-shaped profile
is predominant on the particle side
($\omega>0$).
This asymmetry is due to
the matrix element effect:
$L_{\bbox k}^-$ vanishes at $(\pi/2,\pi/2)$.

In any case of this toy band structure, 
it may be totally 
hopeless to experimentally detect such tiny structures as indicated in
Fig.~8(a).
We see in the following that the realistic band structure of BSCCO
 drastically  changes this situation. 

\subsubsection{The case of real band structure}
Next, we take account of 
$\tilde t_2=-0.550$ and $\tilde t_3=0.087$
to reproduce the
real band structure of
BSCCO
measured by angle-resolved photo emission spectroscopy.\cite{Norman98}
 In Fig.~9(a), we show the profile of
 $N({\bbox  r},\omega)/x \alpha^2$
 at the same points as in  Fig.~8(a).
In this case, inside the overall V-shaped profile
with the sharp peaks at
$\tilde\omega\equiv\omega/\chi_0 \tilde J=\pm 0.323$ associated with the 
$d$-wave
superconducting
gap,
 there appears prominent SPD structure
 around $\tilde\omega=\pm 0.179$ and $\tilde\omega=0.224$.
  In Fig.~9(b), we show the same profile as in Fig.~9(a)
over an wider energy window.
The peaks at $\tilde\omega=\pm 0.79$ are ascribed to the
van-Hove singularity at $(0,\pm\pi)$ and $(\pm\pi,0)$ 
points.\cite{Walter-Wen00}

 The SPD structure inside the V-shaped profile again comes from
 resonant scattering between the fermions with
${\bbox k}$ and ${\bbox k}+{\bbox Q}$.
As seen in Fig.~9(c),
the low energy elliptic contours in the case without $t_2$ and $t_3$ [Fig.~8(b)]
bend  around the $d$-wave nodes [bending of the Dirac cone].
Consequently,
as $\omega$ increases from zero, the
energy contours
$E_{{\bbox k}}=\omega$ and $E_{{\bbox k}+{\bbox Q}}=\omega$ first touch
on the reduced zone boundary at $\tilde\omega=\pm 0.179$
as indicated in Fig.~9(d) and resonance occurs.
Then, at $\tilde\omega=\pm 0.22$ they touch
again at $(\pi/2,\pi/2)$ as indicated in Fig.~9(e) and the second resonance 
occurs.
The reason why
the second resonance comes up only in the electron ($\omega>0$)
side is again ascribed to
the matrix element effect as in the case of $t_2=t_3=0$.
We can say that {\it due to the real band structure
[bending of the Dirac cones around the $d$-wave nodes]
the staggered structure in the LDOS profile becomes far more prominent as 
compared with
the case of $t_2=t_3=0$.}

We see that the SPD structure 
due to the period doubling occurs only
inside the V-shaped profile [see Fig.~9(b)].
In fact, the energy scale at which the SPD structure appears
depends on the band structure parameters [$a_0\propto x$, 
$\Delta_0/\chi_0$, $t_2$, and $t_3$]. 
For a reasonable choice of the parameters
 in the underdoped regime,  however,
the resonance always occur at 
the energy scales below that of the superconducting gap, i.e.,
the SPD structure always appear  inside the V-shaped profile.

%%%%%%%%%%%%%%%%%%%%
%Fig.9 %
\begin{figure}
%\epsfxsize=3.5in
%\centerline{\epsffile{Fig09.eps}}
\smallskip
\caption{
(a) LDOS profile in the case of the real band structure of 
BSCCO, obtained by the perturbative analysis at 
the points
A, B, C, and D indicated in the inset.
The LDOS at B and D are just $\bar N_0(\omega)$.
The staggered   structure around 
$\tilde\omega=\pm 0.179$ and $\tilde\omega=0.224$ comes from
 resonant scattering between the fermions with
${\bbox k}$ and ${\bbox k}+{\bbox Q}$, caused by the period doubling.
The small wiggles outside the V-shaped profile come from numerical
fluctuations.
(b)   The profile 
over a wider energy window  than that of (a).
The peaks at $\tilde\omega=\pm 0.79$ are ascribed to the
van-Hove singularity at $(0,\pm\pi)$ and $(\pm\pi,0)$ points.
(c) The one-particle energy contour
around the $d$-wave node.
 The energy contours $E_{\k}=\omega$ and
$E_{{\k}+{\bigQ}}=\omega$ touch at
 $\tilde\omega=\pm0.179$ and $\pm0.224$ as indicated
in (d) and (e), respectively.
}
%\label{short-fig1}
\end{figure}
%                 %
%%%%%%%%%%%%%%%%%%%

To see a qualitative feature of the doping dependence, 
in Fig.~10(a),
we show the LDOS profile
for $a_0=0.03 \chi_0 \tilde J$ and
 $\Delta_0/\chi_0=0.35$, corresponding to the case of
 a lower doping as compared with
 the case of  $a_0=0.05 \chi_0 \tilde J$ and
 $\Delta_0/\chi_0=0.2$.
 We see that the SPD structure remains robust, although 
 the resonance occurs only once at $\tilde \omega=0.2$.
 Smearing out of the second resonance is due to change of the
 geometry of the Dirac cone around the $d$-wave nodes.
 The shape of the low-energy contours
  change upon changing $\Delta$ as clearly seen by comparing  Fig.~10(b)
  with Fig.~9(c).
  The contours 
  $E_{{\bbox k}}=\omega$ and $E_{{\bbox k}+{\bbox Q}}=\omega$ touch
on the reduced zone boundary only at $\tilde\omega=\pm 0.2$ [Fig.~10(c)].
  As already mentioned, however, location of the $d$-wave nodes
 is independent of $\Delta_0$ and  always shift from $(\pm\pi/2,\pm\pi/2)$
 for finite $\tilde t_3$, i.e, 
 the resonance at $(\pm\pi/2,\pm\pi/2)$ occurs
at the enregy
\be
\tilde \omega^\ast=\pm(2\tilde t_3+a_0/\chi_0 \tilde J).\label{resonance}
\en
 In this respect,
{\it the next nearest neighbor hopping $\tilde t_3$ plays a crucial role
to push the energy scales of the SPD structure toward
visibly finite energy scales.}

%%%%%%%%%%%%%%%%%%%%
%Fig.10 %
\begin{figure}
%\epsfxsize=3.2in
%\centerline{\epsffile{Fig10.eps}}
\smallskip
\caption{
(a) LDOS profile for $a_0=0.03 \chi_0 \tilde J$ and
 $\Delta_0/\chi_0=0.35$, corresponding to the lower doping as compared with
 the case of  $a_0=0.05 \chi_0 \tilde J$.
(b) The one-particle energy contour
around the $d$-wave node.
 The energy contours $E_{\k}=\omega$ and
$E_{{\k}+{\bigQ}}=\omega$ touch at
 $\tilde\omega=\pm0.2$ as indicated
in (c).
}
%\label{short-fig1}
\end{figure}
%                 %
%%%%%%%%%%%%%%%%%%% 

In the perturbative picture presented here, 
the period-doubled perturbation processes form
the \lq\lq mirror image\rq\rq  of the energy bands
with respect to the reduced zone boundary [Fig.~7(a)].
The energy levels $E_{\k}$ and its mirror image $E_{\k+\bigQ}$
cross on the zone boundary
$|k_x|+|k_y|=\pi$, which causes the resonant scattering
at the corresponding energy $\omega^\ast$ [Fig.~7(b)].
It is naturally expected  that
if we go beyond the perturbative scheme
the level crossing would 
be lifted and eventually the period doubling may
cause gap opening in the fermion excitation spectrum as indicated in Fig.~7(c).
This point is confirmed through the exact diagonalization under
uniform $\bv$ approximation as shown below.

\subsection{Exact diagonalization using the uniform $\bv$ approximation}

Next we consider the case of
uniform $\bb{v}$-field:
${\bv}_0=
(v_{0x},v_{0y})$ which may locally
capture the effects of  the circulating $\bv(\r)$.
From Eqs.~(\ref{modchiDelta}), 
(\ref{outsidechi}), and (\ref{outsideDelta}), 
we see that the uniform $\bv_0$ yields
\be
\bar \chi_{ij}=\tilde \chi_{ij}&=& A\cos[\Phi_0+
{(-1)^{i_x+j_y}}(\r_i-\r_j)\cdot{\bv}_0
],\\
\bar \Delta_{ij}=\tilde \Delta_{ij}&=&A\sin[\Phi_0+
{(-1)^{i_x+j_y}}
(\r_i-\r_j)\cdot{\bv}_0
].
\en
An  advantage of the uniform $\bb{v}$ approximation is that
we can exactly diagonalize the corresponding fermion Hamiltonian,
which can be written as
\be
{H}_0^{\rm F}
={1\over 2}
\sum_{{\bbox k}\in\rm RZ}
{\bbox\Psi}_{{\bbox k}\sigma}^\dagger
{\bf{T}}_{{\bbox k}}
{\bf\Psi}_{{\bbox k}\sigma},
\en
where
$
({\bbox\Psi}_{{\bbox k}\sigma})^T=
((\psi_{{\bbox k}\sigma})^T,(\psi_{{\bbox k}+{\bbox Q}\sigma})^T).
$
The $4\times 4$ matrix ${\bf{T}}_{{\bbox k}}$ is given by
\be
{\bf{T}}_{{\bbox k}}
=\pmatrix{
V_{{\bbox k}}+a_{\bbox k}
\tau^3& i W_{{\bbox k}}\cr
i W_{{\bbox k}+{\bbox Q}}&V_{{\bbox k}+{\bbox Q}}+a_{{\bbox k}+{\bbox
Q}}\tau^3},\label{matrixM}
\en
where
$
V_{{\bbox k}}=-\chi_0 \tilde J\tilde\gamma_{\bbox k}\tau^3+\Delta_0
 \tilde
J\tilde\mu_{\bbox
k}\tau^1,
$
and 
$
W_{{\bbox k}}=\Delta_0 \tilde J\tilde\lambda_{\bbox k}\tau^3+\chi_0
\tilde J\tilde\mu_{\bbox k}\tau^1,
$
with
$
\tilde\gamma_{\bbox k}=
\cos v_{0x} \cos k_x+\cos v_{0y} \cos k_y,
\tilde\eta_{\bbox k}       =\cos v_{0x} \cos k_x-\cos v_{0y} \cos k_y,
\tilde\lambda_{\bbox k} =\sin v_{0x} \sin k_x+\sin v_{0y} \sin k_y,
\tilde\mu_{\bbox k}       =\sin v_{0x} \sin k_x-\sin v_{0y} \sin k_y.
$
Noting the fact
that the  field $\bv$ does not modulate
 the hopping  amplitude between the same sublattice sites,
we  take account of the hopping parameters
 $t_2$ and $t_3$ by introducing
\be
a_{\bbox k}=a_0-\tilde t_2 \cos k_x\cos k_y
-\tilde t_3(\cos 2 k_x+\cos 2 k_y).
\en
The one-particle propagator in a $4\times4$ matrix form
is given by
\be
{\bf{G}}^{\rm F}({\k} ,i\omega)=
[i\omega{\bf{1}}-{\bf{T}}_{{\k}}]^{-1},\label{44Green}
\en
where ${\bf{1}}$ denotes a $4\times4$ unit matrix.
As was inferred  from the perturbative analysis,
the unit cell doubling brings about
 the one-particle spectrum split
into two branches in the reduced zone,
$
\pm E_{\bbox k} ^+
$
and
$
\pm E_{\bbox k} ^-
$,
where
\be
E_{\bbox k}^\pm
=[
a_{\bbox k}^2+\gamma_{{\bbox k}}^2+\eta_{{\bbox k}}^2+\lambda_{{\bbox 
k}}^2
+\mu_{\k}^2
\pm 2\{
a_{\bbox k}^2(\gamma_{{\bbox k}}^2+\lambda_{{\bbox k}}^2)+(\eta_{{\bbox
k}}\lambda_{{\bbox k}}
+\gamma_{{\bbox k}}\mu_{{\bbox k}})^2\}^{1/2}
]^{1/2},
\en
with $\k\in{\rm RZ}$. 
To compute the LDOS, we need 11, 33, 13, and 31 components of
${\bf{G}}^{\rm F}({\k} ,i\omega)$ which are explicitly
given in appendix D.

As in the perturbative analysis,
we consider  the LDOS at four distinct points on lattice:
(a) the plaquette-center, (b) the site-top, and (c) the bond-center points.
Repeating an analysis similar to that in appendix C,
we obtain the LDOS in a form
\be
N(\omega)/
x \alpha^2
=
\tilde N_0(\omega)\pm\delta \tilde N(\omega),\label{exactLDOSformula}
\en
where $+$ and $-$ signs alternated plaquette to plaquette,
site to site, or bond to bond for the cases
(a), (b), and (c), respectively.
The uniform counterpart is given exactly the same form as in the case of
the perturbative analysis:
\be
\tilde N_0(\omega)=-{1\over \pi}{\rm Im}\sum_{\k}
M_0(\k)
[{\bf{G}}^{\rm F}(\k,i\omega)]_{11}
\mid_{i\omega\to\omega+i\delta},
\en
where $M_0(\k)$ are given by Eqs.~(\ref{M0a})-(\ref{M0c}).
We here used the relation,
$
[{\bf{G}}^{\rm F}(\k,i\omega)]_{33}=[{\bf{G}}^{\rm F}(\k+\bigQ,i\omega)]_{11},
$
which is explicitly shown in appendix D. 

The staggered counterparts at the plaquette-center and the site-top 
points  are given in a form
\be
\delta \tilde N(\omega)
=-{1\over \pi}{\rm Im}\sum_{\k\in{\rm RZ}}M(\k)
\left([{\bf{G}}^{\rm F}(\k,i\omega)]_{13}
+[{\bf{{G}}}(\k,i\omega)]_{31}
\right)\mid_{i\omega\to\omega+i\delta},
\en
where the matrix elements $M(\k)$ 
are given by
\be
&&M^{\rm plaquette}(\k)=
\sin{k_x}\sin{k_y},\\
&&M^{\rm site}(\k)=1.
\en
As shown in appendix D, we have  the following relation:
$
[{\bf{G}}^{\rm F}(\k,i\omega)]_{13}=
-[{\bf{G}}^{\rm F}(\k,i\omega)]_{31}=
[{\bf{G}}^{\rm F}(\k,i\omega)]_{31}^\ast.
$
Therefore,
$\delta \tilde N(\omega)$ exactly vanishes
at the plaquette-center and the site-top points.

On the other hand, at the bond-center points  we obtain
\be
\delta \tilde N(\omega)=
-{1\over \pi}{\rm Im}
\sum_{\k\in{\rm RZ}}
\sin k_{\mu}
\left(i[{\bf{G}}^{\rm F}(\k,i\omega)]_{13}
-i[{\bf{{G}}}(\k,i\omega)]_{31}\right)
|_{i\omega\to\omega+i\delta},
\en
which remains finite, where
we considered the bond in the $\hat{\bf{e}}_\mu$-direction.
Thus, just as in the perturbative  analysis,
the LDOS  exhibits  staggered pattern
only when measured on the bonds.
Using an explicit form of ${\bf{G}}^{\rm F}$
given in appendix D, we obtain
\be
\tilde N_0(\omega)&=&\sum_{\k}
\cos^2{k_x\over 2}
U_{\k}(\omega)\left[
\delta( \omega-E_{\k}^-)
+
\delta( \omega+E_{\k}^-)
-
\delta( \omega-E_{\k}^+)
-
\delta( \omega+E_{\k}^+)
\right]\label{exactuni}
,\\
\delta \tilde N(\omega)&=&
\sum_{\k\in{\rm RZ}}
\lambda_{x}\tilde U_{\k}(\omega)\sin{k_\mu}
\left[
\delta( \omega-E_{\k}^-)
+
\delta( \omega+E_{\k}^-)
-
\delta( \omega-E_{\k}^+)
-
\delta( \omega+E_{\k}^+)
\right]\label{exactstag}
,
\en
where
$
U_{\k}(\omega)=
[\omega^2+\omega A_{\k}-B_{\k}-C_{\k}/\omega]/2P_{\k},
$
and
$
\tilde U_{\k}(\omega)=-
[
\omega +2 a_{\k} +\tilde C_{\k}/\omega
]/2P_{\k},
$
with $P_{\k}$, $A_{\k}$, ... being given in appendix D. 

\subsubsection{The case of $t_2=t_3=0$}
First, we consider again a toy band structure
with  $t_2=t_3=0$ in
Eq.~(\ref{gamma}).
 In Fig.~11(a),
we show the profile of
 $\tilde N_0(\omega)$
 and $\tilde N_0(\omega)\pm\delta \tilde N(\omega)$ for
 $\bv_0=(0,0.1)$, of which
 direction and strength
locally simulate $\bv(\r)$
around the points B, D, and A,C  in the inset of Fig.~8(a), respectively.
We  used the same parameter set as in the case of Fig.~8(a).
In Fig.~11(b) is indicated 
the energy contour of the lower band $E_{\k}^-$
with  
the corresponding
band structure of $E_{\k}^\pm$
being shown  in Fig.~11(c).
The uniform $\bv_0$ field breaks the original 4-fold symmetry
and the $d$-wave nodes are located slightly off the $\Gamma-$M line.
The van-Hove singularity on the Y-$\Gamma$ line
is caused solely by the superconducting gap
and gives peaks at $\tilde\omega=\pm0.38$.
%\twocolumn[\hsize\textwidth\columnwidth\hsize\csname@twocolumnfalse\endcsname  %%%%%%%%%%%%%%%%%%%%
%Fig. 11%
\begin{figure}
%\epsfxsize=3.5in
%\centerline{\epsffile{Fig11.eps}}
\smallskip
\caption{
(a) Profile
of $\tilde N_0(\omega)$, and
$\tilde N_0(\omega)\pm\delta \tilde N(\omega)$
for the uniform  field $\bv_0=(0,0.1)$
in the case of $t_2=t_3=0$.
In the inset is shown
fine structure
of $\tilde N_0(\omega)$ around $\tilde\omega\sim 0.1$,
detected with higher numerical resolution.
(b) The energy contour of the lower band $E_{\k}^-$
and
(c) the dispersion of $E_{\k}^\pm$
with $\bv_0=(0,0.1)$
along the path $\Gamma(0,0)\to{\rm M}(\pi/2,\pi/2)\to
{\rm Y}(0,\pi)\to\Gamma$. Fine
band splitting
on the reduced zone boundary are magnified
 in the inset.
(d) Profile
of $\tilde N_0(\omega)$, and
$\tilde N_0(\omega)\pm\delta \tilde N(\omega)$
for the uniform  field $\bv_0=(-0.1/\sqrt{2},0.1/\sqrt{2})$.
In the inset is shown
fine structure
of $\tilde N_0(\omega)$ around $\tilde\omega\sim 0.05$,
detected with higher numerical resolution.
(e) The energy contour of the lower band $E_{\k}^-$
and
(f) the dispersion of $E_{\k}^\pm$
with
 $\bv_0=(-0.1/\sqrt{2},0.1/\sqrt{2})$.
 Fine
band splittings
on the reduced zone boundary are magnified
 in the inset.
}
%\label{short-fig1}
\end{figure}
%                 %
%%%%%%%%%%%%%%%%%%%
%]
As expected from the perturbative analysis,
the unit cell doubling causes the gap opening on the
reduced zone boundary between $\tilde\omega=0.10$ and $\tilde\omega=0.12$.
The van-Hove singularities associated
with this gap structure gives rise to
the specific structure in the LDOS profile.
The corresponding fine structure in $\tilde N_0(\omega)$
could be detected with much higher
numerical resolution[720$\times$720 meshes of the Brillouin zone],
as shown in the inset of Fig.11~(a).
The van-Hove singularity at Y$(0,\pi)$ point
is intrinsic to the normal state dispersion $\gamma_k$
and give peaks at $\tilde\omega=\pm0.41$, just as in the case of Fig.~8(a).

We see
 that the staggered modulation profile 
 shown in Fig.~11(a), $\tilde N_0(\omega)\pm\delta \tilde
N(\omega)$,
 is in remarkable agreement
 with Fig.~8(a) obtained by the perturbative analysis.
However, a striking difference is that the dip structure around
$\tilde\omega=0.105$ is now intrinsic to
the modified band structure with van-Hove singularities associated with
the gap opening on the reduced zone boundary
and appears even in the uniform counterpart $\tilde N_0(\omega)$.

 In Fig.~11(d),
we show the LDOS profile  for
 $\bv_0=(-0.1/\sqrt{2},0.1/\sqrt{2})$, 
 where strength of $\bv_0$ is the same 
as in the case of Fig.~8(a), but its
direction
locally simulates $\bv(\r)$
in the 45 degree direction  in Fig.~5.
In Fig.~11(e) is indicated 
the energy contour of the lower band $E_{\k}^-$
with  
the corresponding
band structure of $E_{\k}^\pm$
being shown 
in Fig.~11(f).
The uniform $\bv_0=(-0.1/\sqrt{2},0.1/\sqrt{2})$
 field breaks the original 4-fold symmetry
in a way different from the case of  $\bv_0=(0,0.1)$.
Consequently, the energy scales of the van-Hove singularities responsible for
the dip structure move downward.
The qualitative feature of the profile, however, does not change much
for small magnitude of $\bv_0$ considered here.

\subsubsection{The case of real band structure}
Next, we turn to the real band structure of 
BSCCO.
 In Fig.~12(a),
we show the profile of
 $\tilde N_0(\omega)$
 and $\tilde N_0(\omega)\pm\delta \tilde N(\omega)$ for
 $\bv_0=(0,0.1)$, of which
 direction and strength
locally simulate $\bv(\r)$
around the points B, D, and A,C  in the inset of Fig.~9(a), respectively.
We  used the same parameter set as in the case of Fig.~9(a).
In Fig.~12(b) is indicated 
the energy contour of the lower band $E_{\k}^-$
with  
the corresponding
band structure of $E_{\k}^\pm$
being shown 
in Fig.~12(c).
The van-Hove singularity on the Y-$\Gamma$ line
is caused solely by the superconducting gap
and gives peaks at $\tilde\omega=\pm0.323$.
The uniform $\bv_0$ field breaks the original 4-fold symmetry
and the $d$-wave nodes are located slightly off the $\Gamma-$M line.
The unit cell doubling causes the gap opening on the
reduced zone boundary between $\tilde\omega=0.229$ and 
$\tilde\omega=0.265$,
corresponding to the second resonance in the perturbative analysis 
[the 
second
touch of the energy contour indicated in Fig.9~(d)].
In this case, due to the presence of $t_2$ and $t_3$ additional
van-Hove singularity occurs at $\tilde\omega=0.186$ and $\tilde\omega=0.216$,
corresponding to the first resonance in the perturbative analysis  
[the first touch of the energy contour indicated in Fig.9~(c)].
The corresponding fine structure in $\tilde N_0(\omega)$
could be detected with much higher
numerical resolution[720$\times$720 meshes of the Brillouin zone],
as shown in the inset of Fig.12~(a).
As was mentioned in the perturbative analysis,
the van-Hove singularity at Y$(0,\pi)$ point is pushed
upward as compared with the case of $t_2=t_3=0$ and
lies at the energy $\tilde\omega=\pm0.79$
outside the energy window of Fig.~12(a) just as
in  Fig.~9(b).

%\twocolumn[\hsize\textwidth\columnwidth\hsize\csname@twocolumnfalse\endcsname  
%%%%%%%%%%%%%%%%%%%%
%Fig. 12%
\begin{figure}
%\epsfxsize=3.5in
%\centerline{\epsffile{Fig12.eps}}
\smallskip
\caption{
(a) Profile
of $\tilde N_0(\omega)$, and
$\tilde N_0(\omega)\pm\delta \tilde N(\omega)$
for the uniform  field ${ { \bv}_0}=(0,0.1)$
in the case of the real band structure of 
BSCCO.
In the inset is shown
fine structure
of $\tilde N_0(\omega)$ around $\tilde\omega\sim 0.2$,
detected with higher numerical resolution.
(b) The energy contour of the lower band $E_{\k}^-$
and
(c) the dispersion of $E_{\k}^\pm$
with $\bv_0=(0,0.1)$
along the path $\Gamma(0,0)\to{\rm M}(\pi/2,\pi/2)\to{\rm Y}(0,\pi)\to\Gamma$. 
Fine band splitting
on the reduced zone boundary are magnified
 in the inset.
(d) Profile
of $\tilde N_0(\omega)$, and
$\tilde N_0(\omega)\pm\delta \tilde N(\omega)$
for the uniform  field ${{\bv}_0}=(-0.1/\sqrt{2},0.1/\sqrt{2})$.
In the inset is shown
fine structure
of $\tilde N_0(\omega)$ around $\tilde\omega\sim 0.2$,
detected with higher numerical resolution.
(e) The energy contour of the lower band $E_{\k}^-$
and
(f) the dispersion of $E_{\k}^\pm$ with 
${{ \bv}_0}=(-0.1/\sqrt{2},0.1/\sqrt{2})$. 
%\label{short-fig1}
}
\end{figure}
%                 %
%%%%%%%%%%%%%%%%%%%
%]
We see again
 that the profile in Fig.~12(a),
  in particular the SPD structure inside the V-shaped profile,
 is in remarkable agreement
 with Fig.~9(a) obtained by the perturbative analysis.
However,  as in the case of $t_2=t_3=0$
a striking difference is that the SPD structure is now intrinsic to
the modified band structure with van-Hove singularities associated with
the gap opening on the reduced zone boundary
and appears even in the uniform counterpart $\tilde N_0(\omega)$.
This suggests that in reality
 the SPD structure may be detected not only on the bonds but also at 
sites.

 In Fig.~12(d),
we show the LDOS profile  for
 $\bv_0=(-0.1/\sqrt{2},0.1/\sqrt{2})$, where strength of $\bv_0$ is the same 
as in the case of Fig.~9(a), but
 its
direction
locally simulates $\bv(\r)$
in the 45 degree direction  in Fig.~5.
In Fig.~12(e) is indicated 
the energy contour of the lower band $E_{\k}^-$
with  
the corresponding
band structure of $E_{\k}^\pm$
being shown 
in Fig.~12(f). Qualitative feature of the LDOS profile
given in Figs.~12(a) and 12(d) are quite similar.
Thus, we may say that  
the SPD  structure is robust and 
detectable in all the directions around the vortex center.

We note that in both perturbative and exact analysis,
the SPD structure in the LDOS is predominant on the particle side
($\omega>0$).
We can understand this asymmetry by first turning off the 
superconductivity
and consider the effect of
unit cell doubling.   Since we are doping with holes, the gaps being 
opened
by the unit cell doubling are on the empty side on the Fermi surface. 
Matrix
element effect preserves this particle-hole asymmetry even after we turn 
on
the superconductivity.

\section{Summary and concluding remarks}
In this paper, we have concentrated on
how to  detect a signature of the unit cell doubling
 originated from the SF state
through STM measurement.
Although the signature of the SF state  appears only dynamically in
a uniform SC state, 
a topological defect (vortex) stabilizes
static texture of the boson condensate and the
spatial component of the massless internal gauge field $\ga$.
We determined the texture associated with a single vortex
based on a simple London model.
%The phases of the SU(2) local boson condensate (LBC) are characterized by
%two U(1) angles,
%$\alpha$ and $\phi/2$ associated with 
%U(1)$_{\rm em}$ and U(1)$_{\rm res}$, respectively.
%Both $\alpha$ and $\phi/2$ wind by
%$\pi$ which lead to an appropriate  $hc/2e$ vortex for
%the EM
%gauge field ${\bbox A}({\bbox r})$.
A half flux quantum of the EM gauge field, $\bb{A}$,
penetrates over a huge region $r\alt\lambda_L$, as compared with
a half flux quantum of the internal gauge field, $\bb{a}^3$, which
penetrates  over a region $r\alt\lambda_a$.
Although the fermions do not couple to the EM gauge field, 
they still see
the internal gauge flux tube associated with $\ga$.
By this reason, the topological texture shows up in the 
hopping ($\chi_{ij}$) and pairing ($\Delta_{ij}$) order parameters of the physical electrons and
gives rise to
the staggered modulation of $\chi_{ij}$ and $\Delta_{ij}$
through the gauge invariant 
\lq\lq superfluid velocity\rq\rq
$\bv$ associated with $\ga$ [see Eqs.~(\ref{modchiDelta}) and (\ref{Phiij})].

The most important formula in this paper is Eq.~(\ref{Uij})
which directly tells us that
whereas the center in the vortex core is a SF state, 
as one moves away from the core center, 
a correlated staggered modulation of the hopping amplitude $\tilde\chi_{ij}$
and pairing amplitude $\tilde\Delta_{ij}$
of the {\it physical} electrons  becomes predominant
over the region $\ell_c\alt r\alt\lambda_a$.
Combining the results obtained through the gradient expansion and the
uniform $\bv$ approximation, 
we concluded  that
 the signature of the unit cell doubling may be most prominently detected
through the staggered peak-dip (SPD) structure inside the V-shaped profile
measured on the bonds. 
The real band structure of BSCCO,
in particular
the next nearest neighbor hopping $\tilde t_3$, 
plays a crucial role
to push the energy scales of the SPD structure toward
visibly finite energy scales.
The structure directly originates from
the unit cell doubling, which is stabilized by
the topological texture (phase winding) under 
the external magnetic field. 
In this respect, our effects have little to do with
the $d$-wave symmetry of the superconducting order parameter.
Our finding may be best summarized in Fig.~13. 
%For an illustration in Fig.~13, we assumed $\ell_c=3$ and $\lambda_a=10$.
%
%%%%%%%%%%%%%%%%%%%%
%Fig. 13%
\begin{figure}
%\epsfxsize=3.4in
%\centerline{\epsffile{Fig13.eps}}
\smallskip
\caption{
(a) The best scan path to test our effects is the path
denoted by  \lq\lq scan 1.\rq\rq
(b) The expected LDOS profile measured on the bond-center points
(O sites on a CuO$_2$ plane)
around a single SU(2) vortex with the real band structure of BSCCO being taken
into account.
For an illustration  we assumed $\ell_c=3$ and $\lambda_a=10$.
The staggered peak-dip (SPD) structure appears over  
 $\ell_c \stackrel{<}{\sim} r \stackrel{<}{\sim} 
 \lambda_a$ and vanishes deep inside
 the $d$-wave SC state ($r\gg\lambda_a$).
We expect almost no effects along the scan path denoted by
 \lq\lq scan 2.\rq\rq 
}
%\label{short-fig1}
\end{figure}
%                 %
%%%%%%%%%%%%%%%%%%%
%
The best scan path to test our effects
is shown in Fig.~13(a) as \lq\lq scan 1,\rq\rq
along which the LDOS on the bonds exhibits 
specific peak and dip structure alternating from bond to bond
in a staggered manner as indicated in Fig.~13(b)
 [the LDOS shown here is obtained under the same setting
as in Fig.~9].
The SPD structure appears over the region
$\ell_c\alt r\alt\lambda_a$ and vanishes deep inside
 the $d$-wave SC state.
The core size $\ell_c$ presumably extends
 over a  fermion coherence length
 $\xi_F\sim v_{\rm F}/\Delta_0$ which may amount to
 a few lattice scales\cite{Han-Wang-Lee00} and
the energetics of a single vortex supports that
large value of $\lambda_a/\ell_c$ tends to be favored.
Thus we are hopeful that there is certainly the region
$\ell_c\alt r\alt\lambda_a$ over which our effects are detectable.
Due to the lattice symmetry,
the unit cell doubling effects on the LDOS
is detectable only on the bonds.
Thus, we have just
a typical V-shaped profile of bulk $d$-wave SC 
along the path denoted by \lq\lq scan 2\rq\rq
in Fig.~13(a).
Although the qualitative feature of the LDOS profile may not be so sensitive
to the doping $x$   in the underdoped  regime,
 fine detail of the SPD structure  depends on the doping dependence $x$
 which controls $a_0\propto x$ and $\Delta_0$.
 In particlular, existence or absence  of the second peak/dip depends on $x$.
Nevertheless, we have at least one resonance (or a pair of van-Hove 
singularities) on the reduced zone boundary
at the energy around 
$\omega=\pm(2\chi_0\tilde J\tilde t_3+a_0)$ [Eq.~(\ref{resonance})]
which always lies inside the V-shaped profile  
for reasonable band structure parameters of BSCCO.

As for an experimental setup
for  BSCCO sample, the best place to test our prediction
is 
the O site around the vortex center on the CuO$_2$ plane. 
%In addition, our analysis based on the London model of a single SU(2)
%vortex may become most 
%reliable when the vortex density is very sparse, i.e,
%$H\agt H_{c1}$. 
The size of the Wannier function at the O sites on Cu-O-Cu bonds
is presumably an order of 10$\%$ in magnitude
as compared with the nearest Cu sites.
However, it is noteworthy  that
the STM tunneling into the O sites  
%can
may take place directly via the STM tips,
while
the tunneling into the Cu sites   on the CuO$_2$ 
plane takes place indirectly through
 the Bi atom on the BiO layer.\cite{Zhu00}
Thus, we are hopeful that
the STM signal may more sensitively detect the LDOS profile
at the O sites
than at the Cu sites.
We stress that the 
SPD structure is totally ascribable to the unit cell doubling and the 
robust topological texture.
%That is to say, our effect has little to do with the 4-fold 
%symmetry associated with  the $d$-wave order parameter. 
Therefore, we may safely say that the SPD structure  survives 
any tunneling matrix element effects
and can directly be detected through STM experiment.

\acknowledgements
{
We acknowledge Seamus~Davis,
 Dung-Hai~Lee, Naoto~Nagaosa, Tai-Kai~Ng, Masao~Ogata, and
Subir~Sachdev for  stimulating communication.
J.~K was supported by a Monbusho Grant for overseas research
during his stay at MIT.
P.A.L and X.-G.W acknowledge support by NSF under the MRSEL
Program DMR 98-08491.
X.-G.W. also acknowledges support by NSF grant No. DMR 97-14198.}

%%%%%%%%%%%%%%%%%%%%%%%%%%%%
\appendix
\section{Internal phase of the local boson condensate}
The uniform $d$-wave SC state in the $d$-wave gauge
is described by,
\be
U_{ij}^d&=&-\chi_0\tau^3+(-1)^{i_y+j_y}\Delta_0\tau_1,\\
h_{0i}^d&=&\pmatrix{\sqrt{x}\cr 0},
\en
which is just equivalent to the U(1) MF solution for the $d$-wave SC state.
Now, thanks to
the SU(2) symmetry, the same state can be
described in the SF gauge via the SU(2)
gauge transformation $w_i^\dagger$ given by
Eq.~(\ref{dSFconverter}).
The gauge transformation
converts $U_{ij}^{d}$ and $h_{0i}$ to
\be
U_{ij}^{\rm SF}
&\to&w_i^\dagger U_{ij}^{d} w_j=
-A\tau^3\exp\left[\ii(-1)^{i_x+j_y}\Phi_0\tau^3\right]
,\label{udsFU}
\\
h_{0i}^{\rm SF}&\to&w_i^\dagger h_{0i}^d= \sqrt{x\over 2}
\pmatrix{
1\cr
-\ii(-1)^{i_x+i_y}
}
,\label{udsfbozon}
\en
where
$
A=\sqrt{\chi_0^2+\Delta_0^2},
$
and
$
\Phi_0=\tan^{-1}(\Delta_0/\chi_0).
$
Now, the low energy excitations around the SC state
 in the SF gauge are obtained by fixing
$U_{ij}^{\rm SF}$ and then rotating
the boson condensate in the internal SU(2) space.
The direction in the internal SU(2) space is
specified by the internal angles $\phi$ and $\theta$ as
in Eq.~(\ref{CP1}).
We obtain
this parameterization  more directly through transforming
$h_{0i}^{d}$ by
\be
g_i^\dagger=\exp\left[-{\ii(-1)^{i_x+i_y}{\theta_i\over 2}\tau^1}\right]
\exp\left[-{\ii{\phi_i\over 2}\tau^3}\right],
\en
which we will encounter in Sec.~III~A [see Eq.~(\ref{eg})]
when we make a gauge transformation.

%%%%%%%%%%%%%%%%%%%
%%%%%%%%%%%%%%%%%%%
\section{London model analysis
of a single SU(2) vortex}

Here we apply 
the  London model prescription\cite{Fetter-Hohenberg69}
to a single SU(2) vortex.
Plugging Eq.~(\ref{CP1rep}) into Eq.~(\ref{FK}) gives
$
F_{\rm K}=F_{V}+F_{v},
$
where
\be
F_{V}&=&{x\over 2m_b}\int d\r{\bb{V}}(\r)^2,\label{Femv}\\
F_{v}&=&{x\over 2m_b}\int d\r{\bb{v}}(\r)^2,\label{Fgaugev}
\en
with
\be
\bb{V}
&=&{1\over 2}\bb{\nabla}\varphi_2-{e\over c}\bb{A}
=\bb{\nabla}\alpha-{e\over c}\bb{A}
,\\
\bb{v}&=&{1\over 2}\bb{\nabla}\varphi_2-\bb{a}^3
={1\over 2}\bb{\nabla}\phi-\bb{a}^3
,\label{va}
\en
being the superfluid velocities associated with $\bb{A}$
and $\bb{a}^3$
fields, respectively.
This decomposition
indicates that $\bb{A}$ and $\bb{a}^3$
gauge fields are decoupled at the mean field level.
The stationality condition with respect to $\bb{A}$,
$
\delta(F_V+F_A)/\delta \bb{A}=0,
$
is reduced to 
$
\bb{\nabla}^2{\bb H}(r)-\lambda_L^{-2}\bb{H}(\r)=
-{\phi_0^{\rm em}\over\lambda_L^2}
\hat{\bf{e}}_z
\delta(r),
$
which gives the solution
\be
\bb{H}(r)
=
{\phi_0^{\rm em}\over 2\pi\lambda_L^2}
\hat{\bf{e}}_z K_0\left(r\over \lambda_L\right),
\en
where $K_0$ is the zero-order modified Bessel
function of an imaginary argument.
The London penetration depth is defined as
$
\lambda_L^2=m_bc^2/4\pi e^2x.
$
The EM unit flux is $\phi_0^{\rm em}=hc/2e$ where we retreived Plank constant.
The physical supercurrent associated with $\bb{A}$ becomes
\be
\bb{J}(\r)={c\over4\pi}\bb{\nabla}\times{\bb{H}}(r)={\phi_0^{\rm em}c
\over 8\pi^2\lambda_L^3}
\hat{\bf{e}}_\phi K_1\left(r\over \lambda_L\right),\label{gcu}
\en
which globally circulate around the vortex center
over the length scale $\lambda_L$.

Taking account of Eq.~(\ref{Fareal}), 
the stationality condition with respect to $\bb{a}^3$,
$
\delta(F_v+F_a)/\delta \bb{a}^3=0,
$
is reduced to
$
-2xt\bb{v}(\r)
+\sigma\int d\r'{\bb{\nabla}_r}\kappa(\r-\r')\times\bb{h}(\r')=0.
$
Taking curl of this equation and going to Fourier space,
we obtain
\be
{\bb{h}}(\bb{q})
%&=&\int d\r e^{i\bb{q}\cdot\r}\bb{\nabla}\times \bb{a}^3\non\\
=\hat {\bf{e}}_z{\phi_0^{\rm gauge}\over 1+\lambda_a\kappa_{\bb{q}} \bb{q}^2}
=\hat{\bf{e}}_z{\phi_0^{\rm gauge}\over 1+\lambda_a |\bb{q}| },\label{sola}
\en
where we made use of Eq.~(\ref{ftk}).
The gauge field penetration depth and the unit flux associated with it are
given by
$
\lambda_a=m_b\sigma/x,
$
and
$\phi_0^{\rm gauge}=h/2$,
respectively.
Fouier transform of Eq.~(\ref{sola}) gives
\be
\bb{h}(r)\!=\!{\phi_0^{\rm gauge}\over\lambda_a^2}
\hat{\bf{e}}_z\!
\left[
{\lambda_a\over r}\!-\!{\pi\over 2}
\left\{
\!{\bf H}_0\!\left({r\over\lambda_a}\right)
\!-\!{\bf N}_0\!\left({r\over\lambda_a}\right)
\right\}
\right],\label{gauge-field-strength}
\en
where ${\bf H}_0(z)$ and ${\bf N}_0(z)$ denote the Struve 
function\cite{Struve}
and  the Bessel function of the second kind, respectively.
We  note that
$
\lambda_a/\lambda_L\sim{1\over \sqrt{x}}{e\over c}\ll 1,
$
where we used $\sigma=\sqrt{\tilde J\Delta}\sim 1/\sqrt{m_b}$
($\lambda_L$ 
reaches $\sim 500$ with lattice unit 
in BSCCO).
The essential point here is that we can reasonably assume
the EM gauge field $\bb{H}$ extends much broader than the internal gauge
field $\bb{h}$.
Under this circumstance, the effect of the EM  gauge potential $\bb{A}$ is
negligible as compared with the internal gauge
potential $\bb{a}^3$.
That is to say, in our vortex model,
a half flux quantum of the EM gauge field, $\bb{A}$,
penetrates over a huge region $r\alt\lambda_L$, as compared with
a half flux quantum of the internal gauge field, $\bb{a}^3$, which
penetrates  over a region $r\alt\lambda_a$.
In Fig.~14, 
we show spatial decay 
of $H(r)/(\phi_0^{\rm em}/2\pi\lambda_L^2)$ 
and $h(r)/(\phi_0^{\rm gauge}/\lambda_a^2)$
assuming 
$\lambda_L/\lambda_a=50$ and $\lambda_a=10$ with lattice unit.
The apparent divergence of $h(r)$ at $r=0$ 
should be taken as an artifact of continuum limit,
since there is natural cutoff of an order of the inverse lattice scale.

%%%%%%%%%%%%%%%%%%%%
%\noindent Fig.14
\begin{figure}
%\epsfxsize=3.0in
%\centerline{\epsffile{Fig14.eps}}
\smallskip
\caption{
Spatial dependence of
$H(r)/(\phi_0^{\rm em}/2\pi\lambda_L^2)$ 
and $h(r)/(\phi_0^{\rm gauge}/\lambda_a^2)$
assuming $\lambda_L/\lambda_a=50$ and $\lambda_a=10$ with lattice unit.
Apparent divergence of $h(r)$ at $r=0$ 
should be taken as an artifact of continuum limit.
}
\end{figure}
%                 %
%%%%%%%%%%%%%%%%%%%

We shall now argue how 
 $\lambda_a$ can become larger than $\ell_c$
in terms of energetics of a single vortex.
The energy associated with a single vortex
consists of the following contribution:
the cost for the SF core formation $\epsilon_{\rm core}$,
the electromagnetic contribution  
 $\epsilon_{\rm EM}=F_{V}+F_{A}$,
 and
the contribution of the internal gauge field
  $\epsilon_{\rm gauge}=F_v+F_a$.
The enegy cost for the SF core  formation   is estimated as
as\cite{Lee-Wen00}
$
\epsilon_{\rm core}\sim \sqrt{\tilde J\Delta} \ell_c^2 x^{3/2},
$
which favors   smaller $\ell_c$.
On the other hand, the core size cannot be smaller than $x^{-1/2}$
without costing too much kinetic energy.
Thus, we conclude that the SF core occupies a radius of
$
\ell_c\sim x^{-1/2}
$
at MF level.
In the present scheme, it is quite reasonable to expect that
as the doping $x$ decreases,
the core size becomes larger  because
the energy difference between the SC and the SF state
 decreases as $x\to0$.

The electromagnetic contribution comes from
 $\epsilon_{\rm EM}=F_{V}+F_{A}$ which
reduces to\cite{Fetter-Hohenberg69}
\be
\epsilon_{\rm EM}&=&
{1\over 8\pi}\int_{r>\ell_c} d\r
\left[
{\bb H}^2+\lambda_L^2(\bb{\nabla}\times{\bb H})^2
\right]\non\\
&=&{\pi x\over 4m_b}\ln{\lambda_L\over \ell_c}.
\en
To compute $\epsilon_{\rm gauge}$, we first
take curl of Eq.~(\ref{va}) and go to Fourier space to obtain
\be
\bb{v}(\bb{q})
=-i\pi{\bb{q}\times\hat{\bf{e}}_z\over q^2}
{\lambda_a q\over1+\lambda_a q},\label{vq}
\en
and similarly
\be
\bb{a}^3(\bb{q})=-i\pi{\bb{q}\times\hat{\bf{e}}_z\over q^2}
{1\over 1+\lambda_a q}.\label{FTa3}
\en
%which lead to
%\be
%F_v&=&{x\over 2m_b}\sum_{\bb{q}}|\bb{v}(\bb{q})|^2
%={\pi^2x\over 2m_b}
%\sum_{\bb{q}}{\lambda_a^2\over(1+\lambda_a q)^2},\\
%F_a&=&{\sigma\over 2}
%\sum_{\bb{q}}{1\over q}|q_y a^3_x(\bb{q})-q_x a^3_y(\bb{q})|^2\non\\
%&=&{\pi^2\sigma\over 2}
%\sum_{\bb{q}}{1\over q}
%{1\over(1+\lambda_a q)^2}.
%\en
Recalling $x/2m_b=\sigma/2\lambda_a$, we have
\be
\epsilon_{\rm gauge}
={\pi^2\sigma\over 2}
\sum_{\bb{q}}{1\over q}
{1\over 1+\lambda_a q}
={\pi \sigma \over 2\lambda_a}\ln{\lambda_a\over \ell_c}
.\label{enegauge}
\en
We thus have  
the energy cost associated with a  single vortex:
\be
\epsilon_{\rm vortex}&=& 
\epsilon_{\rm core}+\epsilon_{\rm gauge}+\epsilon_{\rm EM}\non\\
&\sim&
\sqrt{\tilde J\Delta} \ell_c^2 x^{3/2}
+{\sigma\pi\over 2\lambda_a}\ln{\lambda_a\over \ell_c}
+{\pi x\over 4m_b}\ln{\lambda_L\over \ell_c}
.
\en
This result is consistent with a
little bit more qualitative discussion,\cite{Lee-Wen00} i.e.,
a standard $hc/2e$ vortex is possible with the SF core which does not cost
too much energy as $x\to 0$.
The SF core size
$\ell_c$ would like to be small as
possible with the lower bound $\ell_c\sim x^{-1/2}$,
while the size of the gauge field distribution $\lambda_a=m_b\sigma/x$
would like to be
large. 
%%%%%%%%%%%%%%%%%%%
\section{Computation of LDOS}
\subsection{Derivation of Eq.~(\ref{LDOSformula})}
We start with  Eq.~(\ref{startingformula}).
The LDOS at the plaquette-center, site-top, and bond-center points,
as indicated in Fig.~6,
are
given by
\be
N^{\rm plaquette}({\bbox r}_{\rm C},\omega)
&=&-{2x \alpha_{\rm C}^2\over \pi}{\rm
Im}\sum_{i,j=1,2,4,5}
[{\cal G}_{ij}^{\rm F}(\ii\omega)]_{11}
\mid_{\ii\omega=\omega+\ii\delta},\\
N^{\rm site}({\bbox r}_1,\omega)
&=&-{2x \over \pi}{\rm Im}
[{\cal G}_{11}^{\rm F}(\ii\omega)]_{11}
\mid_{\ii\omega=\omega+\ii\delta},\\
N^{\rm bond}({\bbox r}_{\rm B},\omega)
&=&-{2x \alpha_{\rm B}^2\over \pi}{\rm
Im}\sum_{i,j=1,2}
[{\cal G}_{ij}^{\rm F}(\ii\omega)]_{11}
\mid_{\ii\omega=\omega+\ii\delta},\label{abondDOS}
\en
respectively.
The envelope function 
is simulated by
$\alpha_i(\r)=e^{-|\r-\r_i|/\xi}$ ($\xi=1/2$ with lattice unit)
in the
bond direction.
Tuning
$\alpha_i(\r_{\bb{i}})=1$ at the site-top points,
we 
put
$
\alpha_1({\bbox r}_{\rm C}
)=
\alpha_2({\bbox r}_{\rm C})=
\alpha_4({\bbox r}_{\rm C})=
\alpha_5({\bbox r}_{\rm C})
\sim e^{-1} 
\equiv
\alpha_{\rm C}
,
$
and
$
\alpha_1({\bbox r}_{\rm B})=
\alpha_2({\bbox r}_{\rm B})\equiv\alpha_{\rm B}
.
$
Using Eq.~(\ref{wfonsublattice}),
we obtain
\be
\psi_{1}+\psi_{2}+\psi_{4}+\psi_{5}&=&
{4\over\sqrt{2}}
\sum_{\k\in \rm RZ}e^{\ii\k\cdot\r_{\rm C}}
\left[\cos{k_x\over2}\cos{k_y\over2}\psi_{\k}\pm
\sin{k_x\over2}\sin{k_y\over2}\psi_{\k+\bb{Q}}\right],\label{center}\\
\psi_{1}&=&
{1\over\sqrt{2}}
\sum_{\k\in \rm RZ}e^{\ii\k\cdot\r_1}
\left[\psi_{\k}\pm\psi_{\k+\bb{Q}}
\right],\\
\psi_{1}+\psi_{2}&=&
{2\over\sqrt{2}}
\sum_{\k\in \rm RZ}e^{\ii\k\cdot\r_{\rm B}}
\left[\cos{k_x\over2}\psi_{\k}\pm
i\sin{k_x\over2}\psi_{\k+\bb{Q}}\right],\label{bond}
\en
where $-$ and $+$ signs being for the cases where the site 1 belongs to
the A and B sites, respectively. Constructing the propagators with 
these
wave functions, we  obtain 
a formula Eq.~(\ref{LDOSformula}) in a concrete form.
For example, in Eq.~(\ref{abondDOS}) we obtain
\be
\sum_{i,j=1,2}{\cal  G}_{ij}^{\rm F}
={\cal  G}^{\rm F}(\r_{\rm B}, \ii\omega)
\pm\delta
{\cal  G}^{\rm F}(\r_{\rm B}, \ii\omega).
\en
The uniform and staggered counterparts are computed as
\be
&&
{\cal  G}^{\rm F}(\r_{\rm B}, \ii\omega)=
\sum_{\k\in\rm RZ}
\sum_{\k'\in \rm RZ}
e^{i(\k-\k')\cdot \r_{\rm B}}
\left[
 \cos{k_x\over2}\cos{k_x'\over2}{\cal G}^{\rm F}(\k,\k',i\omega)
+\sin{k_x\over2}\sin{k_x'\over2}{\cal G}^{\rm
F}(\k+\bb{Q},\k'+\bb{Q},i\omega)\right]\non\\
&&=
\sum_{\k}
\cos^2{k_x\over2}{\cal G}_0^{\rm F}(\k,\ii\omega),
\en
and
\be
&&\delta
{\cal  G}^{\rm F}(\r_{\rm B}, i\omega)
=i\sum_{\k\in\rm RZ}
\sum_{\k'\in \rm RZ}
e^{i(\k-\k')\cdot \r_{\rm B}}\left[
\cos{k_x\over2}\sin{k_x'\over2}{\cal G}^{\rm
F}(\k,\k'+\bb{Q},i\omega)
-\sin{k_x\over2}\cos{k_x'\over2}{\cal G}^{\rm F}(\k+\bb{Q},\k',i\omega)
\right]\non\\
%&=&
%\sum_{\k\in\rm RZ}\sum_{\k'\in \rm RZ}\sin[(\k-\k')\cdot \r_{\rm B}]
%\sin{k_x\over2}\cos{k_x'\over2}
%\non\\&&\,\,\,\,\,\,\,\,\,\,\,\,\,\,\,\,\,\,\,\,\,\,\,\,\,\,\,\,\,\,\,\,\,\,
%\times\left[
%{\cal G}^{\rm F}(\k',\k+\bigQ,i\omega)+{\cal G}^{\rm F}(\k+\bigQ,\k',i\omega)
%\right]
%\non
%\\
&=&
\sum_{\k\in\rm RZ}
\sum_{\q\sim \rm small}
\sin(\q\cdot \r_{\rm B})
\sin{k_x+{q_x\over2}\over2}\cos{k_x-{q_x\over 2}\over2}
\left[
{\cal G}^{\rm F}(
\k-{\q\over 2},
\k+{\q\over2}+\bigQ,i\omega)
+
{\cal G}^{\rm F}(\k+{\q\over2}+\bigQ,
\k-{\q\over 2},i\omega)
\right],\label{staggB}
\en
which give
Eq.~(\ref{formulauniform})
and
Eq.~(\ref{formulastagg}),
respectively.
At the last step 
we replaced $\k-\k'$ and $(\k+\k')/2$ with $\q$ and $\k$, respectively.

\subsection{Computation of the generalized propagators}

Next we compute
\be
{\cal G}^{\rm F}(\k+{\q\over2}+\bigQ,\k-{\q\over2},\tau)
&=&
-\langle T_\tau
\psi_{\k+{\q\over2}+\bigQ}(\tau)
\psi_{\k-{\q\over2}}^{\dagger}
\rangle,\\
{\cal G}^{\rm F}(\k-{\q\over2},\k+{\q\over2}+\bigQ,\tau)
&=&
-\langle T_\tau\psi_{\k-{\q\over2}}(\tau)
\psi_{\k+{\q\over2}+\bigQ}^{\dagger}
\rangle,
\en
at Born level.
First, we rewrite
the
 perturbation term Eq.~(\ref{perturbation}) in momentum space:
\be
&&\delta H^{\rm F}=\\
&=&
-i\tilde J
\sum_{\k\in \rm RZ}
\sum_{\k'\in \rm RZ}
\sin\left({k_x+k_x'\over2}\right)v_x (\k-\k')
\left[
\psi_{\k}^{\dagger}(\Delta_0\tau^3+\chi_0\tau^1)\psi_{\k'+\bigQ}
-\psi_{\k+\bigQ}^{\dagger}(\Delta_0\tau^3+\chi_0\tau^1)\psi_{\k'}
\right]
\non\\
&+&\ii J
\sum_{\k\in \rm RZ}
\sum_{\k'\in \rm RZ}
\sin\left({k_y+k_y'\over2}\right)v_y (\k-\k')
\left[
\psi_{\k}^{\dagger}(\Delta_0\tau^3-\chi_0\tau^1)\psi_{\k'+\bigQ}-
\psi_{\k+\bigQ}^{\dagger}(\Delta_0\tau^3-\chi_0\tau^1)\psi_{\k'}
\right].\non
\en
By replacing 
$\k-\k'$ and $(\k+\k')/2$ with $\q$ and $\k$, respectively,
and recalling Eq~(\ref{vq}), we reach Eq.~(\ref{conpert}).
The first order contribution of $\delta H^{\rm F}$
to the propagator is obtained as
\be
{\cal G}^{\rm F}(\k+{\q\over2}+\bigQ,\k-{\q\over2},\tau)
=-\sum_{\omega}e^{\ii\omega\tau}
{\cal G}_{0}^{\rm F}(\k+{\q\over2}+\bigQ,\ii\omega)
{\cal C}_{\k+\bigQ}(\q){\cal G}_{0}^{\rm F}(\k-{\q\over2},\ii\omega),
\en
i.e.,
\be
{\cal G}^{\rm F}(\k+{\q\over2}+\bigQ,\k-{\q\over2},\ii\omega)
=
-
{\cal G}_{0}^{\rm F}(\k+{\q\over2}+\bigQ,\ii\omega)
{\cal C}_{\k}(\q){\cal G}_{0}^{\rm F}(\k-{\q\over2},\ii\omega),
\en
Similarly, we obtain
\be
{\cal G}^{\rm F}
(\k-{\q\over2}
,\k+{\q\over2}+\bigQ,\tau)
=
-{\cal G}_{0}^{\rm F}(\k-{\q\over2},\ii\omega)
{\cal C}_{\k}(\q){\cal G}_{0}^{\rm F}(\k+{\q\over2}+\bigQ,\ii\omega).
\en

Recalling Eq.~(\ref{freepropagator}),
we explicitly write down perturbative corrections:
\be
&&
{\cal G}^{\rm F}(\k-{\q\over 2},\k+{\q\over2}+\bigQ,i\omega)
+
{\cal G}^{\rm F}(\k+{\q\over2}+\bigQ,
\k-{\q\over 2},i\omega)
\non
\\
&=&-
{1\over \ii\omega-E_+}
{1\over \ii\omega-E_-}
\left[
U_+
{\cal C}_{\k}(\q)
U_-
+
U_-
{\cal C}_{\k}(\q)
U_+
\right]\non\\
&-&
{1\over \ii\omega+E_+}
{1\over \ii\omega+E_-}
\left[
V_+
{\cal C}_{\k}(\q)
V_-
+
V_-
{\cal C}_{\k}(\q)
V_+
\right]
\non\\
&-&
{1\over \ii\omega-E_+}
{1\over \ii\omega+E_-}
\left[
U_+
{\cal C}_{\k}(\q)
V_-
+
V_-
{\cal C}_{\k}(\q)
U_+
\right]\non\\
&-&
{1\over \ii\omega+E_+}
{1\over \ii\omega-E_+}
\left[
V_+
{\cal C}_{\k}(\q)
U_+
+
U_-
{\cal C}_{\k}(\q)
V_+
\right],\label{explicit}
\en
where
\be
E_+&=&E_{\k+{\q\over2}+\bigQ},\,\,\,\,\,\,\,
E_-=E_{\k-{\q\over2}},\\
U_\pm&=&
{1\over2}\left[1+{\gamma_\pm\tau^3+\eta_\pm\tau^1}\right]
,\,\,\,\non\\
V_\pm&=&
{1\over2}\left[1-{\gamma_\pm\tau^3-\eta_\pm\tau^1}\right],
\en
with
$
\gamma_+={\gamma_{{\bbox k}+\q/2+{\bbox Q}}
/ E_{{\bbox k}+\q/2+{\bbox Q}}},
$
$
\gamma_-={\gamma_{\k-\q/2}/ E_{\k-\q/2}},
$
$
\eta_+={\eta_{{\bbox k}+\q/2+{\bbox Q}}/ E_{{\bbox k}+\q/2+{\bbox Q}}},
$
and
$
\eta_-={\eta_{\k-\q/2}/ E_{\k-\q/2}}.
$
Taking 11 component of Eq.~(\ref{explicit}) and then performing
analytic continuation, $i\omega\to\omega+i\delta$, we
reach Eq.~(\ref{perturbationresult}).

\section{Explicit form of ${\bf{G}}^{\rm F}$}

By simply taking inverse of the matrix 
$i\omega{\bf{1}}-{\bf{T}}_{{\k}}$
with ${\bf{T}}_{{\k}}$
given by Eq.~(\ref{matrixM}),
we obtain an explicit form of ${\bf{G}}^{\rm F}({\k} ,i\omega)$. 
The 11 and 33 components are given by
\be
&&[{\bf{G}}^{\rm F}(\k,i\omega)]_{11}=[{\bf{G}}(\k+\bigQ,i\omega)]_{33}\non\\
&=&{
(i \omega)^3+A_{\k}(i\omega)^2
-B_{\k}i\omega -C_{\k}\over D(\k,i\omega)
}\\
&=&
{U_{1\k}\over i\omega-E_{\k}^-}+
{V_{1\k}\over i\omega+E_{\k}^-}+
{U_{2\k}\over i\omega-E_{\k}^+}+
{V_{2\k}\over i\omega+E_{\k}^+},
\en
where
$D(\k,i\omega)=\det[i\omega-{\bf{T}}_{{\k}}]=
(i\omega-E_{\k}^-)
(i\omega+E_{\k}^-)
(i\omega-E_{\k}^+)
(i\omega+E_{\k}^+),
$
and
\be
A_{\k}&=&a_{\k}-\gamma_{\k},\\
B_{\k}&=&(a_{\k}+\gamma_{\k})^2
+\eta_{\k}^2+\lambda_{\k}^2+\mu_{\k}^2,\\
C_{\k}&=&
(a_{\k}+\gamma_{\k})(a_{\k}^2-\gamma_{\k}^2-
\lambda_{\k}^2+\mu_{\k}^2)
+
(a_{\k}-\gamma_{\k})\eta_{\k}^2
+2\eta_{\k}\lambda_{\k}\mu_{\k}.
\en
The generalized coherence factors are given by
\be
U_{1\k}&=&{1\over 2}{1\over P_{\k}}
\left[
(E_{\k}^-)^2+E_{\k}^- A_{\k}-B_{1\k}-C_{1\k}/E_{\k}^-
\right],\\
V_{1\k}&=&{1\over 2}{1\over P_{\k}}
\left[
(E_{\k}^-)^2-E_{\k}^- A_{\k}-B_{\k}+C_{\k}/E_{\k}^-
\right],\\
U_{2\k}&=&{1\over 2}{1\over P_{\k}}
\left[
-(E_{\k}^+)^2-E_{\k}^+ A_{\k}+B_{\k}+C_{\k}/E_{\k}^+
\right],\\
V_{2\k}&=&{1\over 2}{1\over P_{\k}}
\left[
-(E_{\k}^+)^2+E_{\k}^+ A_{\k}+B_{\k}-C_{\k}/E_{\k}^+
\right]
,
\en
where
\be
P_{\k}\equiv (E_{\k}^-)^2-(E_{\k}^+)^2
=-4\sqrt{
a_{\k}^2(\gamma_{\k}^2+\lambda_{\k}^2)+(\eta_{\k}\lambda_{\k}
+\gamma_{\k}\mu_{\k})^2}.
\en
Similarly,
\be
&&
[{\bf{G}}^{\rm F}(\k,i\omega)]_{13}=
[{\bf{G}}^{\rm F}(\k,i\omega)]_{31}^\ast\non\\
&=&-i \lambda_{\k}{
(i\omega)^2
+2a_{\k}i\omega + \tilde C_{\k}
\over D(\k,i\omega)}\\
&=&-i \lambda_{\k}\left[
{\tilde U_{1\k}\over i\omega-E_{\k}^-}+
{\tilde V_{1\k}\over i\omega+E_{\k}^-}+
{\tilde U_{2\k}\over i\omega-E_{\k}^+}+
{\tilde V_{2\k}\over i\omega+E_{\k}^+}\right],\non
\en
where
\be
\tilde C_{\k}=
a_{\k}^2-\gamma_{\k}^2-\lambda_{\k}^2+\eta_{\k}^2-\mu_{\k}^2
+2\gamma_{\k}\eta_{\k}\mu_{\k}/\lambda_{\k} ,
\en
\be
\tilde U_{1\k}&=&-{1\over 2}{1\over P_{\k}}
\left[
E_{\k}^- +2a_{\k}+\tilde C_{\k}/E_{\k}^-
\right],\\
\tilde V_{1\k}&=&-{1\over 2}{1\over P_{\k}}
\left[
-E_{\k}^- +2a_{\k}-\tilde C_{\k}/E_{\k}^-
\right],\\
\tilde U_{2\k}&=&-{1\over 2}{1\over P_{\k}}
\left[
-E_{\k}^+ -2a_{\k}-\tilde C_{\k}/E_{\k}^+
\right],\\
\tilde V_{2\k}&=&-{1\over 2}{1\over P_{\k}}
\left[
E_{\k}^+ -2a_{\k}+\tilde C_{\k}/E_{\k}^+
\right].
\en

Now,
LDOS at the midpoint on the bond
connecting $\r_i$ and $\r_i+\hat{\bf{e}}_\mu$
 is given in the form of Eq.~(\ref{exactLDOSformula})
 with
\be
\tilde N_0(\omega)&=&
\sum_{\k}
\cos^2{k_{\mu}\over 2}
\left[
U_{1\k}\delta( \omega-E_{\k}^-)
+
V_{1\k}\delta( \omega+E_{\k}^-)
+
U_{2\k}\delta( \omega-E_{\k}^+)
+
V_{2\k}\delta( \omega+E_{\k}^+)
\right]
,\\
\delta \tilde N(\omega)&=&
\sum_{\k\in{\rm RZ}}\lambda_{\k}\sin k_\mu
\left[
\tilde U_{1\k}\delta( \omega-E_{\k}^-)
+
\tilde V_{1\k}\delta( \omega+E_{\k}^-)
+
\tilde U_{2\k}\delta( \omega-E_{\k}^+)
+
\tilde V_{2\k}\delta( \omega+E_{\k}^+)
\right].
\en
These equations  further reduce to  Eqs.~(\ref{exactuni})
and (\ref{exactstag}).

%\end{multicols}

\begin{references}
\bibitem{comment-on-gauge}
When we write Eq.~(\ref{sfan}) as
$U_{ij}^{\rm SF}
=\pmatrix{-{\chi_{ij}^{\rm SF}}^\ast&0
\cr 0&\chi_{ij}^{\rm SF}},$
an additional gauge transformation
$\chi_{ij}^{\rm SF}\to i\chi_{ij}^{\rm SF}$
and ${\chi_{ij}^{\rm SF}}^\ast\to -i{\chi_{ij}^{\rm SF}}^\ast$
leads to an alternative gauge choice:
$
\hat U_{ij}^{\rm SF}=-i \chi_0-(-1)^{i_x+j_y}\tau^3\Delta_0, 
$
which was used in the papers by 
Lee-Nagaosa-Ng-Wen,\cite{Lee-Nagaosa-Ng-Wen98}
and Lee-Wen.\cite{Lee-Wen00}
This gauge transformation just causes $(\pi/2,\pi/2)$ shift of the
Borillouin zone and leads to the $b_1$ and $b_2$ boson bands 
having the same minimum at $(\pi/2,\pi/2)$.
Then, the local boson condensate is written as
$
{\hat h_{i}}^{\rm SF}
=\sqrt{x}e^{-i(i_x+i_y)\pi/2}
\pmatrix{z_{i1}\cr z_{i2}},
$
instead of Eq.~(\ref{CP1}).
%Due to the different gauge choice, the texture of the 
%$\bbox{I}$-vector shown in FIG.~1(a) is different from that
%given in FIG.~3 of Lee-Wen's paper.\cite{Lee-Wen00}
This change of the gauge choice
is of course just apparent and does not change
any physical consequence of our vortex model.
\bibitem{comment1} To have some feeling of this situation, we consider
a non-interacting electrons  on a linear chain
described by a toy Hamiltonian
$H=-\sum_{i}
[t+(-1)^i\delta][c_{i+1}^\dagger e^{i(-1)^{r_i}\phi}c_i+{\rm H.c.}]$.
It is easy to see that the staggered orbital currents flow on bonds,
$j_{ij}=it(c_{j\sigma}^\dagger 
c_{i\sigma}-c_{i\sigma}^\dagger c_{j\sigma})$,
if $\delta=0$ and $\phi\neq0$, while
the LDOS on the bonds may become staggering if
$\delta\neq0$ and $\phi=0$.
The orbital current flowing and the staggered L
DOS are complementary to each other.
\end{references}
\end{document}